%% file: main.tex
\documentclass[conference]{IEEEtran}
\IEEEoverridecommandlockouts

\usepackage{cite}
\usepackage{amsmath,amssymb,amsfonts}
\usepackage{algorithmic}
\usepackage{graphicx}
\usepackage{subcaption}
\usepackage{textcomp}
\usepackage{xcolor}
\usepackage[hyphens]{url}
\usepackage{fancyhdr}
\usepackage{hyperref}
\usepackage[normalem]{ulem} 
\usepackage{booktabs}
\usepackage{physics}
\usepackage{qcircuit}
\usepackage{wrapfig}
\usepackage{multirow}
 
\makeatletter
\newcommand{\linebreakand}{%
  \end{@IEEEauthorhalign}
  \hfill\mbox{}\par
  \mbox{}\hfill\begin{@IEEEauthorhalign}
}
\makeatother

\begin{document}

\title{Loss-correcting fault-tolerant quantum computing architecture for neutral atoms\\
\thanks{\IEEEauthorrefmark{1}Equal Contribution.\\}
\thanks{\IEEEauthorrefmark{2}Corresponding author.\\}


}

\author{
\IEEEauthorblockN{Sanaa Sharma\IEEEauthorrefmark{1}}
\IEEEauthorblockA{
\textit{University of Cambridge}\\
Cambridge, UK\\
ss3241@cam.ac.uk
}
\and
\IEEEauthorblockN{Yutaka Hirano\IEEEauthorrefmark{1}}
\IEEEauthorblockA{
\textit{Nanofiber Quantum Technologies}\\
Tokyo, Japan\\
yutaka.hirano@nano-qt.com
}
\and
\IEEEauthorblockN{Akihisa Goban}
\IEEEauthorblockA{
\textit{Nanofiber Quantum Technologies}\\
Tokyo, Japan\\
akihisa.goban@nano-qt.com
}
\and

\linebreakand

\IEEEauthorblockN{Hayata Yamasaki}
\IEEEauthorblockA{
\textit{Nanofiber Quantum Technologies}\\
Tokyo, Japan\\
\textit{The University of Tokyo}\\
Tokyo, Japan\\
hayata.yamasaki@gmail.com
}
\and
\IEEEauthorblockN{Shinichi Sunami\IEEEauthorrefmark{2}}
\IEEEauthorblockA{
\textit{Nanofiber Quantum Technologies}\\
Tokyo, Japan\\
\textit{University of Oxford}\\
Oxford, United Kingdom\\
shinichi.sunami@nano-qt.com
}

\and
\IEEEauthorblockN{Prakash Murali\IEEEauthorrefmark{2}}
\IEEEauthorblockA{
\textit{University of Cambridge}\\
Cambridge, UK\\
pm830@cam.ac.uk
}

}

\maketitle
\renewcommand{\figureautorefname}{Fig.}
\renewcommand{\tableautorefname}{Table}
\renewcommand{\equationautorefname}{Eq.}
\renewcommand{\sectionautorefname}{Sec.}
\renewcommand{\subsectionautorefname}{Sec.}
\renewcommand{\subsubsectionautorefname}{Sec.}
\renewcommand{\appendixautorefname}{Appendix}

\begin{abstract}
Neutral-atom arrays are one of the leading qubit technologies for building large-scale, fault-tolerant quantum computers (FTQC). A dominant source of error on this platform is qubit loss, which accrues with every qubit movement and gate operation.  The existence of such an error undermines the promises of existing architectural work. First, standard error-correction routines target stochastic Pauli errors and cannot correct loss errors; consequently, most FTQC performance analyses are not directly compatible with the presence of dominant loss errors. Second, compilation and routing decisions have a major impact on the overall loss magnitude. Existing architectures and compilation frameworks optimize against Pauli-error cost models and are often loss-agnostic, so such a schedule may, in fact, increase exposure to the loss channel, and the resulting resource estimates may not reflect the true cost of fault-tolerant operation.

In this work, we comprehensively model the effect of qubit loss on neutral-atom FTQC and develop a loss-tolerant transversal-gate architecture.
We treat qubit loss not merely as a predetermined error parameter but as a dynamic budget spent across a whole program, which allows us to control it by co-designing layout, compilation, and decoding. 
In particular, we target the physical implementation without the need for separate storage and entangling zones, which eliminates the repeated SLM-AOD handoffs and long-distance shuttling that dominate loss in other layouts. 
We also develop compiler optimizations that maximize gate parallelism while respecting the RF tone budget and AOD bandwidth constraints. 
Our work couples these with a loss-aware, delayed-erasure decoder and an end-to-end loss-aware magic state cultivation protocol. 
Overall, our work improves accumulated loss per syndrome-extraction round by 1.25× on average (up to 2.15×) and CX routing time by up to 8.5× against a zoned baseline, and reduces logical error rates by over two orders of magnitude compared to current architectures. 
Our framework also informs concrete device targets, such as continuous reloading rates, AOD counts, and shuttling trajectory choices. 
We expect these insights to be important for system architects as neutral-atom hardware scales.
\end{abstract}

\input{sections/1-introduction}

\input{sections/2-background}

\input{sections/3-research_problem}

\input{sections/4-compiler_methods}

\input{sections/5-architectural_modelling}

\input{sections/6-experimental_setup}
\input{sections/7-evaluation}

\input{sections/8-discussion}

\bibliographystyle{IEEEtran}
\bibliography{refs}

\end{document}

%% file: sections/1-introduction.tex
\section{Introduction}

Quantum computation promises computational speedups for hard problems in domains such as cryptanalysis~\cite{shor1997polynomial-time, gidney2021how} and quantum chemistry~\cite{lee2021even, kim2022fault-tolerant}.
The error rates of existing platforms are in the range of $10^{-4}$--$10^{-3}$ \cite{Evered2023} per operation \cite{Bluvstein:2025ped}, whereas the error rates required by practical applications range from $10^{-12}$ to $10^{-6}$ \cite{beverland2022assessing} per operation.
Fault-tolerant quantum computation (FTQC) is necessary for realizing large-scale quantum computers.
Neutral atom quantum computers have recently emerged as one of the leading platforms for implementing FTQC, offering a large number of qubits with high homogeneity, all-to-all connectivity, long coherence times, and high-fidelity entangling gates~\cite{Bluvstein2024, Henriet2020quantumcomputing,Manetsch2024ATA, Sunami2025}. There have been recent demonstrations of several primitives of FTQC, such as repeated error correction~\cite{computing2026quantumerrorcorrectiontoric}, logical circuit execution with error detection~\cite{Bluvstein2024}, and magic state distillation~\cite{rodriguez2025experimental}.

\begin{figure*}[t]
    \centering
    \includegraphics[width=0.99\linewidth]{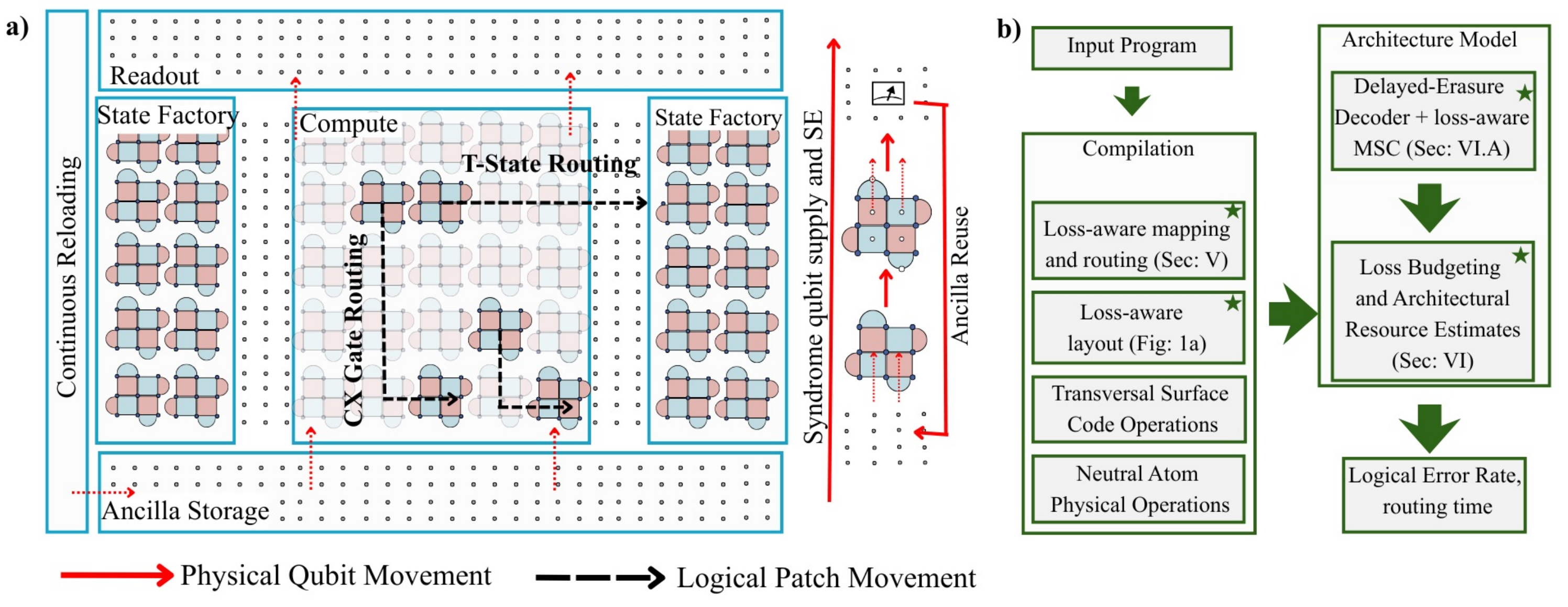}
    \caption{a) Loss-aware layout: all logical operations happen inside the compute zone, where logical qubits are shuttled across other qubits for transversal CXs/ T-state routing. Ancilla qubits are measured in readout zone, and reused following a reinitialization in storage zone. b) Loss-aware layout modeling and compilation framework. Starred boxes: this work's contribution.    }
    \label{fig:overall-layout}
\end{figure*}

A major issue of the neutral atom platform is that it is particularly susceptible to qubit loss. Qubit loss accounts for a large fraction of the errors in important physical operations such as 
entangling gates~\cite{Evered2023}, qubit movement~\cite{Bluvstein2021AQP, Hwang:25}, readout~\cite{graham2023midcircuit, 66s8-jj18}, as well as idle phases~\cite{grimm2000optical, Manetsch2024ATA}.
Loss errors are particularly challenging to detect or correct in conventional error correction protocols and standard syndrome extraction techniques \cite{baranes2026leveraging,suchara2015leakage,perrin2025quantum}. 
Further, lost qubits must be filled with fresh qubits in order to continue the computation.

Recently, significant progress in mitigating qubit loss has been made on both theoretical and experimental fronts. From the theoretical side, loss-correcting protocols have been developed, including a loss-detection unit~\cite{perrin2025quantum, cong2022hardware}, a teleportation-based protocol~\cite{sahay2023biased}, and swap-based syndrome extraction~\cite{suchara2015leakage, baranes2026leveraging}. 
Some of these are assisted by high-performance decoding protocols utilizing the delayed erasure information obtained from the delayed detection of loss events at the time of qubit measurements ~\cite{baranes2026leveraging}.
Experimentally, continuous qubit reloading of fresh neutral atom qubits has been developed to fill the lost sites~\cite{chiu2025continuous, Li2025FastCA, norcia2024iterative}. 
Despite these promising advances, current system designs remain largely oblivious to qubit loss, presenting an opportunity to develop loss-aware compilation and system architecture that is capable of achieving large-scale FTQC with realistic overhead.

To handle loss errors effectively and develop scalable architectures, three challenges must be addressed.
First, existing system-level FTQC architecture studies~\cite{10946298,10.1145/3676642.3736128,j2fw-ccmy} generally ignore loss errors, whereas existing loss-tolerant protocols~\cite{suchara2015leakage, baranes2026leveraging, perrin2025quantum} lack system-level evaluation.
In particular, the overhead of resource-state preparation in the presence of qubit loss has not been evaluated, despite this process being a major bottleneck in FTQC~\cite{gidney2024magic}.
Second, physical qubit layouts must mitigate the substantial transport loss inherent in prevalent zoned architectures~\cite{Bluvstein2024}. In these designs, forcing logical patches to travel long distances to dedicated entangling regions necessitates multiple optical trap transfers and accumulates loss with every gate.   
Third, computation must be compiled and scheduled to minimize the loss events. Existing neutral-atom compilers focus on optimizing gate fidelity, but loss accumulates far more sharply in the FTQC setting because of the additional movement introduced by T-state routing and syndrome extraction. We require compilers that minimize loss while maximizing gate parallelism within the available resources, a joint objective that current work does not address.


To overcome these challenges, we present a full-stack loss-aware FTQC evaluation framework and co-designed compiler and architecture optimizations that mitigate the effects of qubit loss. At the software and modeling level, we introduce a comprehensive simulation pipeline for the surface code incorporating SWAP-based syndrome extraction and delayed-erasure decoding~\cite{baranes2026leveraging}, enabling detailed tracking and correction of loss throughout computation. At the hardware level, we propose a loss-aware layout (\autoref{fig:overall-layout}) based on spatially selective gate execution~\cite{66s8-jj18, Graham2021MultiqubitEA})  that eliminates the massive long-distance shuttling overheads of zoned layouts. To map applications efficiently onto this architecture, we develop a loss-aware compiler that features AOD-configuration-aware routing to maximize gate parallelism within strict RF bandwidth constraints. Further, we introduce a transfer-deferred scheduling strategy that minimizes consecutive pickup-and-dropoff handoffs of the same qubit, significantly curbing transfer loss. We evaluate our framework by running a diverse suite of quantum benchmarks through our compiler to track per-operation loss and couple it with circuit-level delayed-erasure decoder simulations. Our contributions are:
\begin{itemize}
\item We present the first framework that tracks loss end-to-end for neutral-atom FTQC: from per-operation loss contributions on a compiled schedule, through delayed-erasure decoding, to a logical error rate and the code distance it demands. Prior work assumes a circuit-independent loss rate, decoupled from loss accumulation across the architecture~\cite{baranes2026leveraging, perrin2025quantum}, while compilers estimate Pauli fidelity alone~\cite{10.1145/3676642.3736128,10.1109/ISCA59077.2024.00030}, neither yields the code distances that loss-dominated operations demand.
\item 
Our work develops a delayed-erasure-aware soft-output decoder, which we use to give the first
evaluation of fold-transversal surface code
cultivation~\cite{sahay2025fold} under realistic loss.
We demonstrate that loss-aware decoding-assisted cultivation is highly efficient, incurring a spacetime overhead equivalent to the cultivation without loss errors at lower two-qubit error rates.

  \item Overall, our work reduces accumulated loss per syndrome extraction round by $1.25\times$ on average (up to $2.15\times$), and routing time by up to $8.5\times$ for 100-qubit applications, over state-of-the-art zoned architectures~\cite{10946298}. The logical error rate advantage grows with system size, from $2.9\times$ at 32 to $80.6\times$ at 100 qubits on Fermi-Hubbard lattices. 

 \item Our work reexamines a common assumption about zoned neutral-atom architectures. While spatially separating storage and entangling zones works well for near-term devices, in deep FTQC circuits the long-distance shuttling and repeated SLM-to-AOD transfers accumulate atom loss to unsustainable levels, bottlenecking scaling.
 \item We provide targets for device parameters in the presence of atom loss. Importantly, atom reloading rates need to improve by two orders of magnitude from current abilities to meet application requirements.   
\end{itemize}

%% file: sections/2-background.tex
\section{Background}

\subsection{Neutral atom array for quantum computing} \label{subsec:background-netral-atom-array}

Single atoms encode qubits in their internal states, such as hyperfine states, and are trapped in individually addressable optical tweezers, a tightly focused laser beam that confines atoms at its focus through dipole force~\cite{Henriet2020quantumcomputing}.
Atoms can either be held in static tweezers generated by a \textit{spatial light modulator} (SLM) \cite{Kim2016SLM}, which host a large number of trap sites, or in a dynamically reconfigurable tweezer array generated by crossed \textit{acousto-optic deflector} (AOD) pairs with a limited number of tweezers per AODs.
Atoms trapped in optical tweezers can be moved in space by \textit{chirping} (sweeping) the RF tones that are used to generate the optical pattern, resulting in the tweezers being steered in the desired direction~\cite{Guo2025AOL, Bluvstein2024}. 

Two-qubit entangling gates are mediated by the Rydberg blockade effect, where two atoms in close proximity are simultaneously excited to highly excited states with strong dipole-dipole coupling over micron-scale distances, resulting in a controlled phase (CZ) gate~\cite{Saffman2016}.
Two physical implementations of the Rydberg gates are possible: the first is by the broad excitation beam, which illuminates a wide \emph{entangling zone} such that pairs in close proximity interact via CZ gates. This is used in zoned architectures (\autoref{fig:zoned_layout}(a)) and requires qubit reconfiguration to select appropriate qubit pairs for the gate.
The second is implemented by focused Rydberg excitation beams, such that only the spatially addressed qubit pair can interact~\cite{66s8-jj18}, enabling the gate to be executed without qubit reconfiguration, with a limited interaction range of typically up to around 10 \textmu m.
Coherent single-qubit gates are performed by site-resolved Raman coupling laser beams~\cite{Bluvstein2024}.
State readout is performed by spin-selective fluorescence imaging~\cite{graham2023midcircuit}.



\subsection{Fault-tolerant quantum computing with surface code} 
\label{subsec:background-ftqc-surface-code}


An $[[n=d^2, k=1, d]]$  rotated surface code~\cite{fowler2012surface} encodes one logical qubit in a two-dimensional square lattice of $d \times d$ data qubits.
Error correction with the surface code is typically performed by extracting the syndrome using $d^2-1$ ancilla qubits, which interact with the data qubits via CX gates, followed by measurements that provide the syndrome bits, which can be decoded by the minimum-weight perfect matching (MWPM) algorithm~\cite{Higgott2025sparseblossom}.

\begin{figure}[t]
    \centering
    \includegraphics[width=\linewidth]{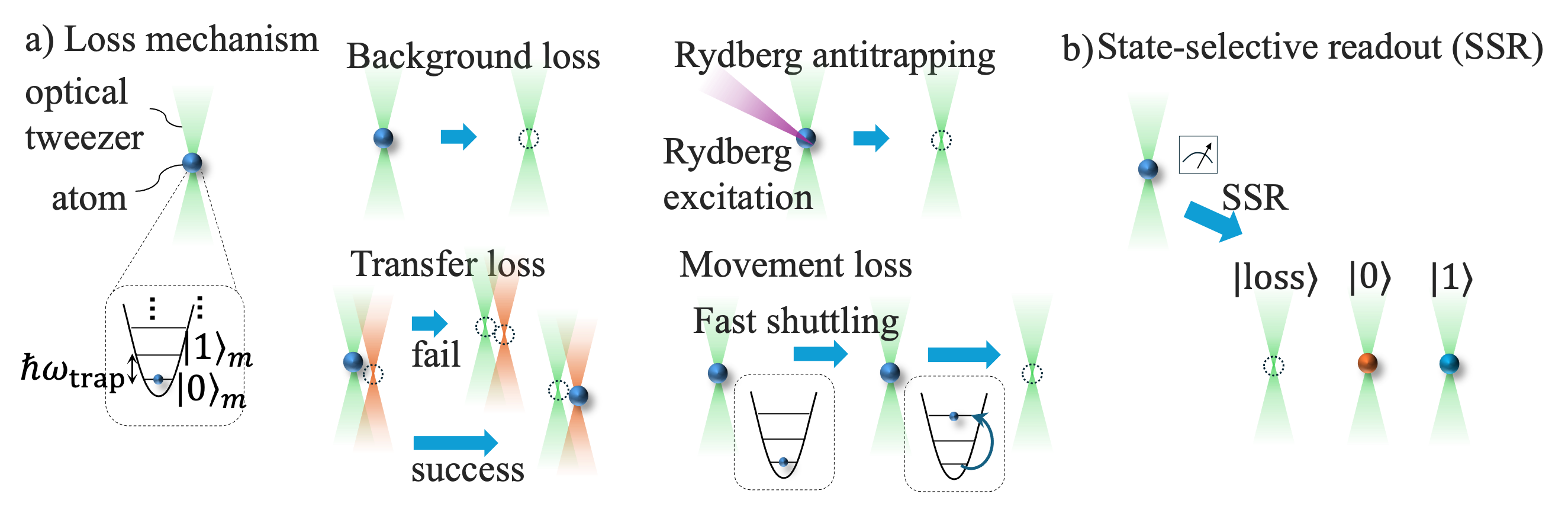}
    \caption{Atom loss mechanisms and state-selective readout.}
    \label{fig:loss_mechanisms}
\end{figure}

Universal quantum computation can be performed by transversal gates, following protocols in \cite{zhou2025low,cain2025fastcorrelateddecodingtransversal} that require only $O(1)$ rounds of syndrome extraction per logical operation.
Transversal CXs are implemented by performing pairwise physical CX gates between the data qubits comprising the logical qubits, while Hadamard gates are implemented by applying physical Hadamard gates to the data qubits and then rotating the lattice of qubits by $\pi/2$~\cite{Chen2026,zhou2025low}.
The $S$ and $T$ gates are implemented by gate teleportation using transversal implementations developed in ~\cite{sunami2025transversalsurfacecodegamepowered}.


%
In particular, $\ket{T}$ resource state is prepared by magic state cultivation (MSC)~\cite{gidney2024magic}; we adopt the variant in Ref.~\cite{sahay2025fold}, which leverages nonlocal connectivity for low spacetime overhead.
MSC consists of three stages: injection, cultivation, and escape.
At the injection stage, a magic state is encoded non-fault-tolerantly.
At the cultivation stage, the state is distilled using error detection and postselection.
At the escape stage, the magic state is re-encoded into a larger QEC code that supports the low logical error rate (LER) of the resulting magic state.
At the end of the escape stage, postselection is performed based on the complementary gap~\cite{gidney2025yoked}.


\subsection{Loss errors in neutral-atom array} \label{subsec:background-loss-errors}
Atom loss is a dominant non-Pauli error on neutral-atom platforms: a qubit leaves the optical trap and becomes unavailable for subsequent operations~\cite{chow2024leakage,baranes2026leveraging}.
A lost qubit voids all subsequent gates involving that qubit, while its intended interaction partners may experience effective single-qubit errors, generating correlated errors that propagate within and across logical blocks~\cite{baranes2026leveraging,Evered2023}.
Unlike Pauli errors, the lost qubit is absent from the codespace rather than perturbed within it, so conventional syndrome extraction cannot diagnose loss.

\textbf{Source of atom losses:}
\label{subsec:background-atom-loss-source}
Fig.~\ref{fig:loss_mechanisms} illustrates atom loss mechanisms, including background-gas collisions that set the vacuum-limited trap lifetime, antitrapping effects during entangling gates, motional heating during tweezer transfer and qubit shuttling, and loss during imaging~\cite{Bluvstein2021AQP,Falconi2025, madjarov2020high}.

\textbf{Loss-detecting qubit measurements:} \label{subsec:background-loss-detecting-measurements}
Qubit loss is detected by \emph{state-selective readout} (SSR), a projective measurement that distinguishes three outcomes $\ket{0}$, $\ket{1}$, and loss~\cite{Bluvstein:2025ped, senoo2025}.
Because SSR is performed at measurement time rather than at the moment of loss, the detection is delayed, requiring a dedicated decoding protocol to correctly incorporate the diagnosis of the loss.
SSR can be realized by using a state-dependent optical trap at the time of the measurement~\cite{Bluvstein:2025ped}, or by leveraging the \emph{shelving} state of the atom for alkaline-earth-like atoms such as Ytterbium~\cite{senoo2025}.

%% file: sections/3-research_problem.tex
\section{Related Work} \label{sec:related-work}
\textbf{Compilers:} Existing neutral-atom compilation frameworks optimize physical-qubit routing across both single-zone arrays and basic two-zone layouts~\cite{10.1109/ISCA59077.2024.00030,    10946298, 10.1145/3676642.3736128}; their NISQ-focused abstractions fail to support FTQC well. 
For example, Atomique~\cite{10.1109/ISCA59077.2024.00030} targets a monolithic array, minimizing SWAP overhead with an AOD router that parallelizes movement, and models a heating-driven channel for transport loss. 
These optimizations do not carry over to fault tolerance where periodic syndrome extraction demands dedicated zones for storage, reloading, etc., which a single processor zone cannot accommodate at scale.
Compilers targeted at zoned architectures~\cite{10946298, 10.1145/3676642.3736128}, shuttle qubits to an entanglement zone for two-qubit gates while shielding idle qubits in a storage zone. This repeated inter-zone
movement is costly in both loss and time. 
Moreover, resource estimates,  such as in
 \cite{zhou2025low, cain2026shor}, use depolarizing
noise models, treating loss either as outside the model or as a fixed,
favorable rescaling of the physical error rate rather than a
circuit-dependent quantity that accumulates with routing.
Across this literature,  
circuit fidelity computations omit atom loss or assume a single loss type. These assumptions are unsuitable for FTQC, where loss converts to erasures that affect both the LER and decoding.

\textbf{Loss correction protocols:} Our delayed-erasure decoder uses the swap-based syndrome extraction circuit introduced in Ref.~\cite{baranes2026leveraging}, which also presents several other syndrome extraction circuits.
For example, the teleportation-based scheme offers higher error thresholds and lower LERs  at the cost of additional spacetime overhead for preparing resource states. The same work also explores mid-circuit erasure conversion that requires additional hardware capabilities.
Ref.~\cite{perrin2025quantum} studies two small circuits, called loss detection units (LDUs), that identify data-qubit loss: standard and teleportation-based LDUs.


\section{Research Problem} \label{sec:research-problem}

\begin{figure}[t]
    \centering
    \includegraphics[width=0.99\linewidth]{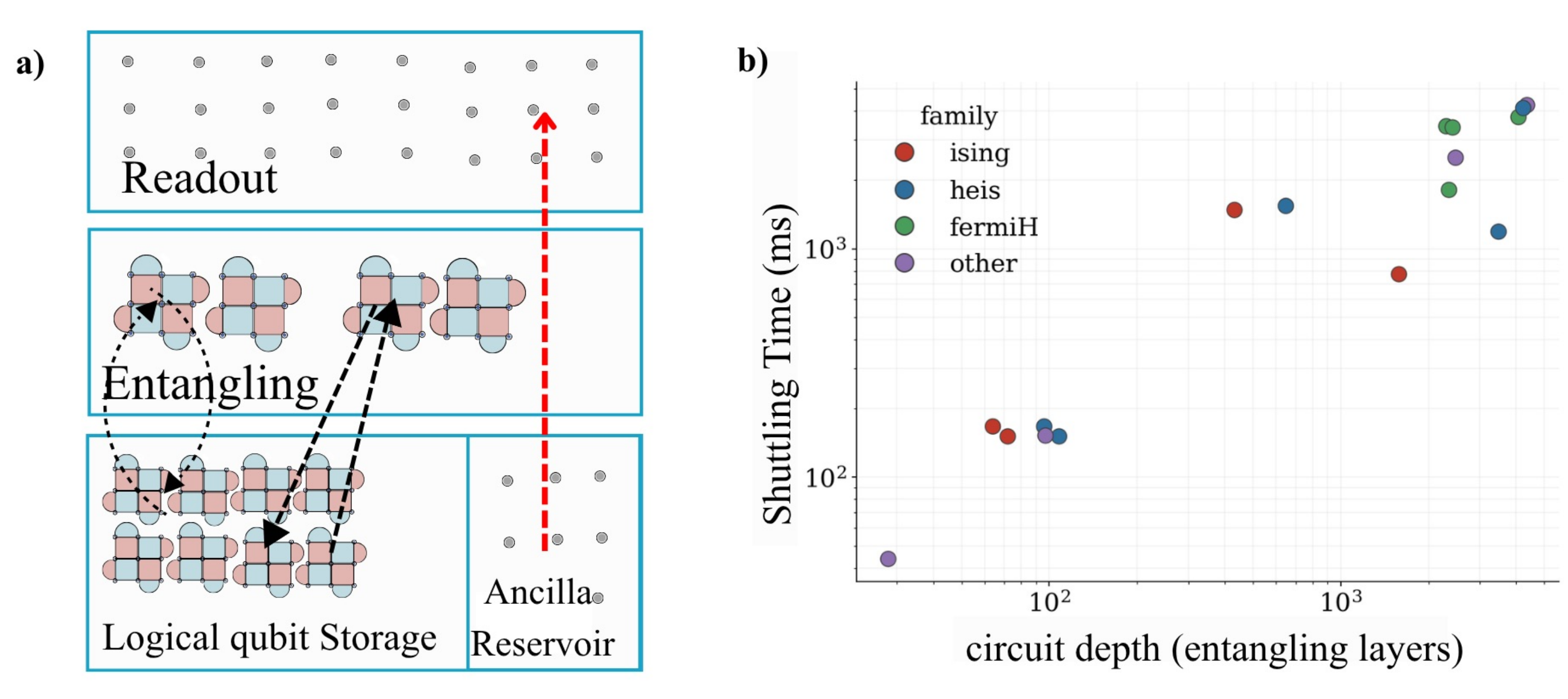}
    \caption{a) Current zoned architecture layout: entangling operations take place in a separate zone, increasing the routing overhead. b) Shuttling time scales with circuit depth.  }
    \label{fig:zoned_layout}
\end{figure}

 \subsection{Inter-zone communication times that worsen with scale}
Atoms shuttle large distances in zoned architectures because of the inter-zonal movement for each operation \autoref{fig:zoned_layout}. Moreover, shuttling also accrues loss based on the distance moved. Atom loss during shuttling is a function of how \textit{far} and how \textit{fast} an atom is moved. A trapped atom suffers from motional heating under finite-time transport \cite{Hwang:25}, and large-distance movements cause  an increase in both execution time and the total loss accumulated under shuttling.

\autoref{fig:zoned_layout}(b) plots the total transport time against circuit depth for the different benchmarks in the zoned architecture. As the number of entangling gates in a circuit grows, the shuttling to and from the entangling zone scales proportionally, driving the cumulative transport distance up sharply. Since transport time $\propto$ distance, this also translates directly into longer execution times. Our work explores whether architectural enhancements such as unifying the entangling and storage zones of current architectures can reduce movement distances and time, without compromising hardware implementation feasibility. 



\subsection{Transfer losses due to inter-zone movement}
\label{sec:transfer_loss}

\begin{figure}[t]
    \centering
    \includegraphics[width=0.99\linewidth]{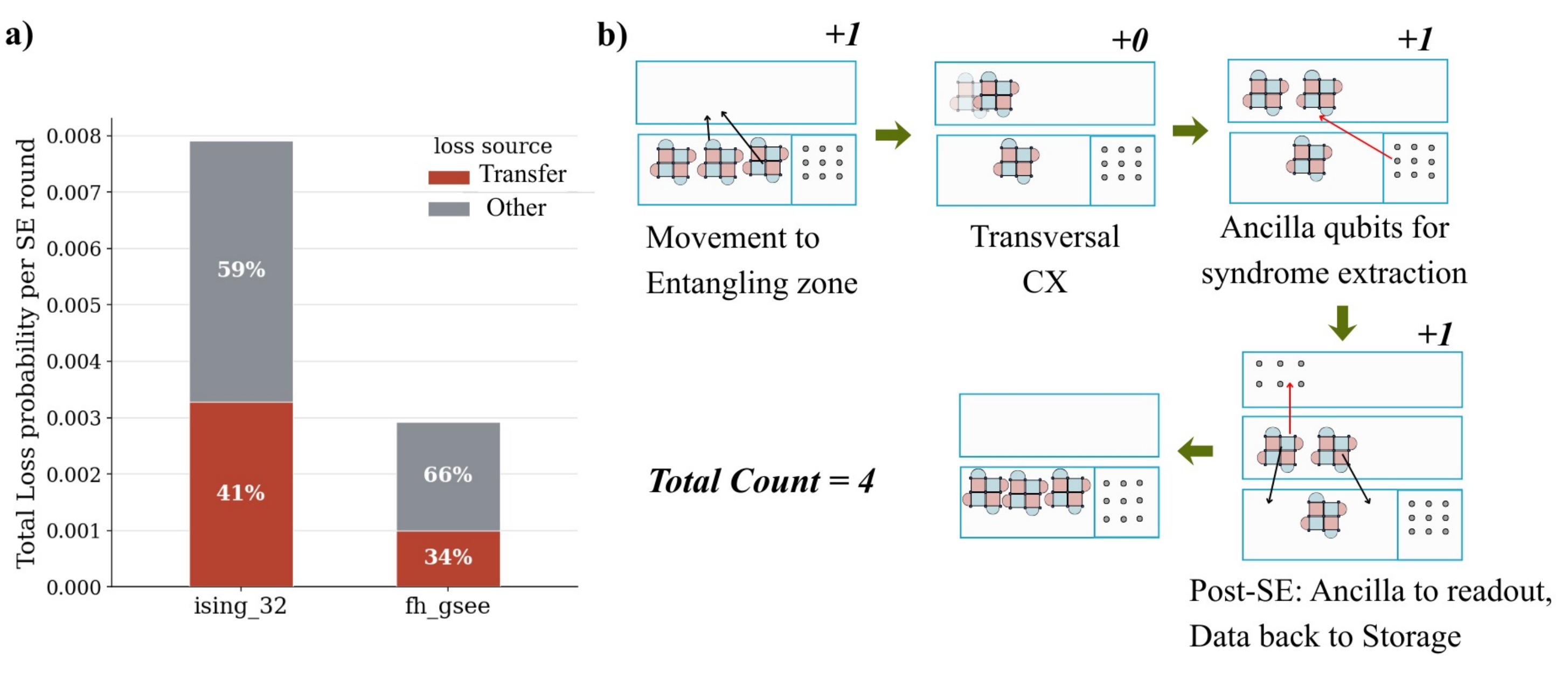}
    \caption{ a) Transfer loss is a major source of loss because of recurrent movement between AOD-SLM traps. b) In zoned layout, transfer loss for each qubit is counted $4\times$ per gate. }
    \label{fig:transfer_loss}
\end{figure}

Transfer losses occur when atoms need to be moved from SLMs to AODs prior to movement, and back to SLMs  at the end of a movement sequence.
\autoref{fig:transfer_loss}(a) quantifies the transfer-loss contribution to total loss due to SLM-AOD transfer for two benchmarks compiled with a zoned-architecture compiler~\cite{10946298}. 
Transfer loss accounts for $\sim40\%$ of the total loss probability accumulated between atom transfers with syndrome-extraction (SE) rounds. The rest of the loss comes from transport, two-qubit gates as well as imaging loss. 
Implementing a gate typically requires, on average, two AOD-SLM transfers (pickups) of the data qubits. This count grows in the zoned architecture of \autoref{fig:transfer_loss}(b), where routing to and from the entangling region, combined with syndrome extraction, doubles it: two transfers to reach the entangling region and two more to return after the SE step. 

\subsection{Loss-awareness in decoding and architecture modelling}
Existing compilation and qubit-routing frameworks for neutral-atom architectures optimize primarily for Pauli errors and do not model atom loss, even though it is a dominant and qualitatively distinct error mechanism on this platform.
Loss-tolerant protocols and decoders have been proposed~\cite{suchara2015leakage,baranes2026leveraging,perrin2025quantum}.
For example, Ref.~\cite{baranes2026leveraging} presents delayed-erasure decoding and swap-based syndrome extraction, both of which we adopt in this work.
Ref.~\cite{perrin2025quantum}, as mentioned in \autoref{sec:related-work},  use LDUs to detect loss.
However, these loss-tolerant protocols have not been integrated with compilation and routing in an end-to-end framework.
An end-to-end loss-aware framework must therefore jointly account for compilation, routing, tailored physical operations, and decoding.

 \textit{Can we unify the storage and entangling zones in the zoned architecture to reduce transfers and minimize pickup losses? Can an alternative to the zoned architecture lower the total shuttling time (and hence the accumulated shuttling loss) while preserving fault-tolerant operation and respecting AOD bandwidth constraints? To what extent can loss-aware decoding improve logical performance, particularly for magic state generation?}

%% file: sections/4-compiler_methods.tex
\section{Compiler Methods}
\label{sec:loss_aware_architecture}
\autoref{fig:overall-layout}(b) illustrates our end-to-end pipeline. The circuit of a target application is first passed to the compiler, which places logical qubits on the layout and routes gates subject to two inputs: the size of the compute zone and the number of AOD. Scheduling is performed on the loss-aware layout, where the AOD budget determines how many gates execute in parallel. The compiled schedule is then passed to the architectural model, which combines the loss realized by the compiler with the delayed-erasure decoder's precomputed loss budget to yield the LER and the total runtime, including routing time.

The loss-aware compilation aims at reducing the loss through two complementary mechanisms. First, we introduce a \textit{loss-aware layout} (\autoref{fig:overall-layout}(a)) that shortens the movements needed for transversal two-qubit gates and T-state routing. Second, it minimizes the number of SLM–AOD transfers: rather than returning an atom to its static (SLM) trap after each gate, atoms participating in consecutive gates remain held in the mobile (AOD) traps, avoiding the loss incurred on every transfer between the traps. To support these excursions in the AOD, the underlying SLM traps along the path are switched off, suppressing crosstalk between both the trap arrays.


    
    

\subsection{Input programs and Target Architecture} 
The pipeline takes a single input circuit, supplied as OpenQASM and normalized to a Clifford+T gate set via gridsynth~\cite{ross2016optimalancillafreecliffordtapproximation}, and uses it in two ways. For the initial mapping, it extracts the circuit's CX gates and groups them into maximal sets of mutually qubit-disjoint gates, capturing the two-qubit dependency structure that placement must respect. Once the qubit placement is fixed, the full circuit drives routing. Atom transport is induced by the two-qubit gates and by the movements to the buffer zone for T gates; the compiler isolates these transport-inducing operations and applies its routing optimization to them, while single-qubit gates (H and Paulis) are implemented in place. The transport operations are scheduled into layers with parallel gates using an as-soon-as-possible (ASAP) policy.

We compile to the fault-tolerant instruction set exposed by a neutral-atom surface code architecture \autoref{table:ISA}\cite{sunami2025transversalsurfacecodegamepowered}. Logical qubits are arranged on a two-dimensional grid, the compute zone \autoref{fig:overall-layout}(a).  Unlike the zoned layout of \autoref{fig:zoned_layout}(a), where qubits travel to a separate zone for each two-qubit operation, here all logical operations occur in the compute zone. Dedicated storage and readout zones serve the ancilla qubits: they move from storage into the compute zone for syndrome extraction, then to readout for non-destructive imaging, after which they return to storage and are reinitialized (\autoref{fig:overall-layout}(a)).

\begin{table}[!t]
\caption{Logical-level instruction Set~\cite{sunami2025transversalsurfacecodegamepowered}}
\label{table:ISA}
\centering

\renewcommand{\arraystretch}{1.4}
\begin{tabular}{|p{0.08\textwidth}|p{0.36\textwidth}|}
\hline
Transport (\texttt{MOVE})
&  The AOD picks up a set of patches from SLM, displaces and releases them to SLM.  A single AOD is able to pick up more than one logical patch and move them provided certain constraints are met (see Sec.~\ref{sec:aod_routing}).
\\ \hline
Logical \texttt{CX} &  overlap two logical patches  ($< r_{rydberg}$ distance) and apply transversal gate via selective Rydberg gate.
\\ \hline
Logical \texttt{H}
&  Applied in place on a patch.
\\ \hline
Logical \texttt{T, S}
&  Gate teleportation consuming a resource state at the dedicated buffer.
\\
\hline
\end{tabular}
\end{table}

\subsection{AOD Configuration-Aware Routing}
\label{sec:aod_routing}

In practice, current architectures operate with no more than two to three AODs \cite{10946298, Manetsch2024ATA}.
The natural unit for AOD-based atom transport is a \textit{crossed pair} (one horizontal, one vertical AOD on the same beam), since each AOD deflects along only one axis.
Adding additional pairs is constrained by factors such as the finite optical aperture of the microscope objective shared by all beam paths and the rapid growth of intermodulation distortion products with tone count \cite{Bluvstein:2025ped}.
The compiler, therefore, treats AOD count as a fixed architectural parameter rather than a free variable, and extracts parallelism by routing within the available AOD budget.

\subsubsection{Initial Mapping} We place application qubits in the compute zone via a simulated-annealing pass minimizing a parallelism cost \texttt{pl\_factor}.It is the ratio of total execution layers (including sublayers from AOD routing constraints) to layers in the unconstrained circuit, where 1 preserves the original gate parallelism and larger values quantify the AOD-imposed slowdown. The annealer starts from a deterministic seed, a single static placement fixed for the whole circuit. The two qubits of each first-layer gate occupy horizontally adjacent sites from $(0,0)$, with remaining qubits filling the grid lexicographically. We select the candidate minimizing the mean  \texttt{pl\_factor}  across all layers.
\subsubsection{Routing Constraints} \label{subsec:routing-constraints}

\begin{figure}[t]
    \centering
    \includegraphics[width=0.99\linewidth]{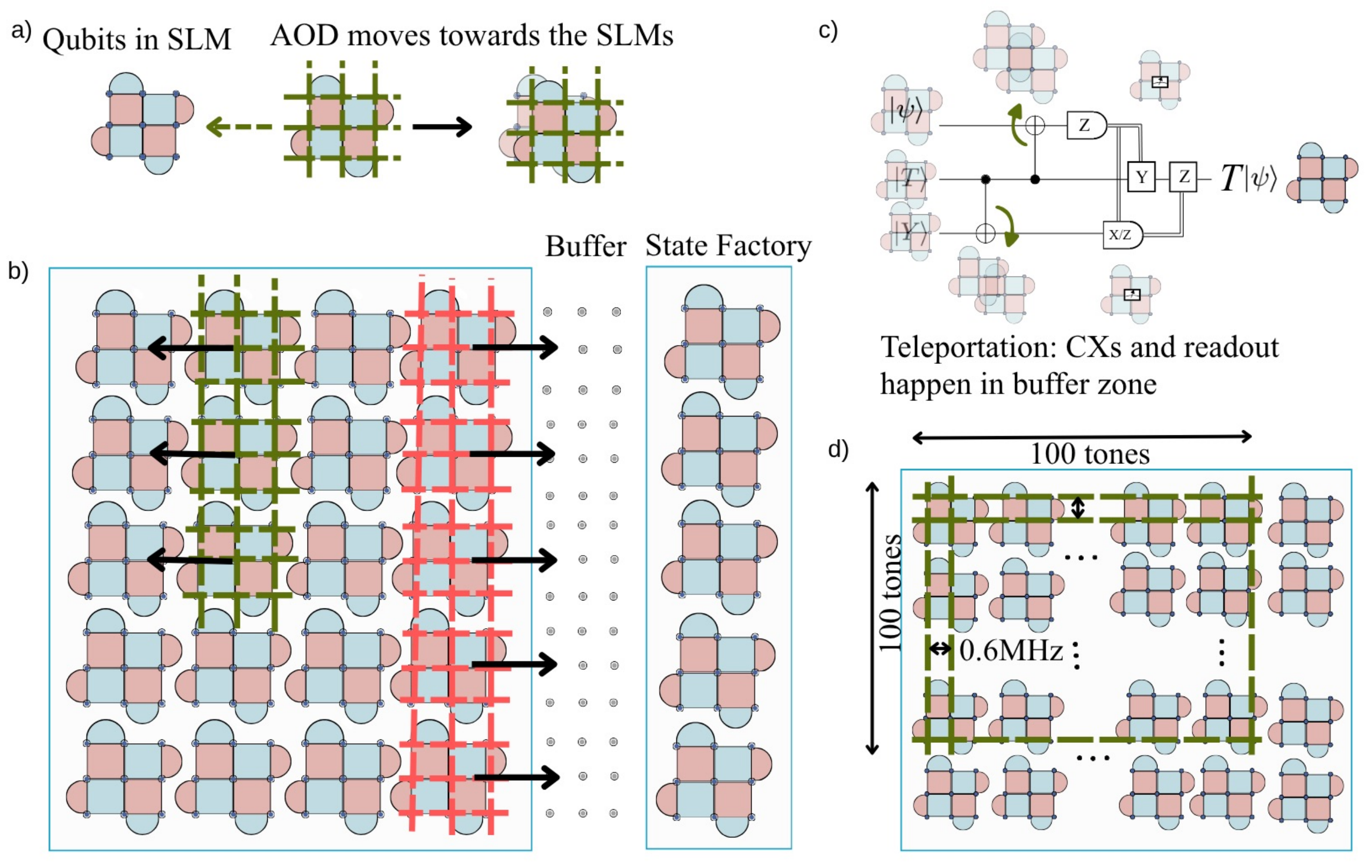}
    \caption{a) Transversal gate implementation in compute zone. b) Two movements facilitated by AODs in compute zone: Transversal gates (green crossed AODs) and T-state routing to buffer zone (red crossed AODs). c) S and T Gate teleportation takes place in buffer zone.  d) Maximum number of RF tones enable parallel gate movements by an AOD. }
    \label{fig:aod_movement}
\end{figure}

There are two movements that should be handled in FTQC compilation for neutral atoms: movement of one logical patch to the other for a CX gate \autoref{fig:aod_movement}(a and b), and movement to the buffer zones for a rotation gate \autoref{fig:aod_movement}(b and c) (T or S gate through gate teleportation \cite{cain2025fastcorrelateddecodingtransversal}).
Both of these implementations require movements of logical patches interleaved between stationary SLM traps. 
The key to parallelism within this budget is that an AOD displaces all of its trapped atoms rigidly by a single common vector (\autoref{fig:aod_movement}(b)). Moves that share a direction can be carried in one AOD load, whereas moves in different directions cannot. \textbf{We maximize parallelism through AOD-configuration-aware routing in the compute zone, which steers data-qubit movements in common directions so that many operations are transported together by a single AOD}. 

What are the gate implementation limits of a single AOD?  A principal bottleneck on how many atoms can be transported together is the
finite RF bandwidth of the AOD \cite{Manetsch2024ATA,endres2016, Bluvstein2024}. In a crossed-pair AOD, the two axes are driven
by two independent crystals, each with its own bandwidth $B$, so the $x$- and
$y$-tones occupy separate spectra and are constrained independently rather than
through their product \autoref{fig:aod_movement}(d). Moving a block of atoms spanning $r$ rows and $c$ columns
costs $r$ tones on one axis and $c$ tones on the other; with a minimum tone
spacing $\Delta f$ (set by cross-talk). Each axis must
satisfy 
\begin{equation}
  N_x \, \Delta f \;\leq\; B, \qquad
  N_y \, \Delta f \;\leq\; B,
  \label{eq:rf-cap}
\end{equation}
where $N_x$ and $N_y$ are the numbers of simultaneous tones on the two axes. We fix $B \sim 60~\mathrm{MHz}$, representative of AODs used in recent
large-scale demonstrations~\cite{Manetsch2024ATA, Bluvstein2024}. The
  $\Delta f$ is set by parametric heating: neighbouring
tweezers beat at their frequency difference, and as the spacing is reduced
this modulation  degrades
lifetimes~\cite{endres2016}. Taking $\Delta f \sim 0.6~\mathrm{MHz}$ gives
a per-axis ceiling of $N_x, N_y \lesssim 100$ tones, so the binding
constraint is the RF bandwidth occupied \emph{on each axis}.

Our compiler enforces Eq.~\eqref{eq:rf-cap} directly during routing. For each transport step it admits only as many atoms as the per-axis tone budget allows, so that the occupied bandwidth on neither axis exceeds $B$. When a layer's tone demand would exceed the available bandwidth on either axis, the surplus moves are reassigned to another AOD. If the AOD budget is likewise exhausted, they are deferred to a subsequent transport step. Each step is thus compiled to a set of atoms that fits within the AOD's RF bandwidth, and any oversized layer is partitioned into successive bandwidth-compliant steps.

To determine the actual routing paths, we use two methods.

\textbf{Simulated Annealing:} To exploit AOD parallelism within each circuit layer, we route all interactions in the layer using a simulated-annealing pass over a discrete space of candidate paths. For every interaction in the layer (CX gates and rotation gates transported to the buffer), we enumerate a bank of candidate paths on the patch grid: \textit{L-} and \textit{Z-shaped} routes between logical patches, on top of the straight paths. The shapes  are chosen to facilitate separable AOD transport that is both axis-aligned and cheap. The annealer state is a vector that assigns each interaction either an index into its candidate bank or a flag deferring it to the next layer. The cost function is a weighted sum of four physically motivated terms: path conflicts, AOD-budget, the number of deferred interactions, and total path length. Weights are chosen to impose a strict priority ordering (collisions$ >$ AOD budget $>$ deferrals$ > $length). AOD reuse is counted strictly: two paths share an AOD only if their inter-cell displacement vectors are identical at every step. 
\label{sec:simulated_annealing_routing}

\textbf{Deterministic Routing:} 
We apply a deterministic alternative as a fallback with the SA described above. Under a tight per-layer AOD budget, the SA cost function can mass-defer. Interactions that do not fit the budget, or that fail to lower the cost, are assigned to such routing schemes that aim to maximize operations in a layer.  
 Between the two routers, the one that packs more operations in a layer is chosen for that layer. First is a \emph{column-grouped buffer router} that sends logical patches to the buffer zones (for T gates). Patches are grouped by compute-zone column, and each group is dispatched to the nearer buffer edge so that a group crossing directly to a buffer row is realized as a single rigid displacement. The crossed-AOD compatibility conditions, together with the tone and aperture budget, then bound how many such displacements can be carried concurrently. Because AODs are assigned first-fit to buffer slots, the order in which the slots are placed determines what remains available to those that follow. The router enumerates these orderings exhaustively, and candidates are ranked lexicographically by the number of distinct displacements. Other ranking factors include deferrals (of operations that do not fit), transit through occupied patch cells, and total path length breaking successive ties. The ordering with the best score (lowest cost) is then selected.

The second method uses \emph{Hungarian Minimum-Cost Matching}, applying the same per-layer interaction framing in two stages. First, for each interaction in the layer (CX gates and $T$ rotations), it enumerates all simple grid paths between the endpoints within a small detour budget above the Manhattan distance (default 2 grid steps), discarding self-overlapping paths. Each interaction's candidate set is trimmed to at most three paths by a cell-popularity score.
The survivors are converted into a labeling structure: every pairwise intersection of two interactions' candidate paths becomes a shared conflict label, while non-overlapping paths receive unique free labels, capturing exactly which paths compete for the same physical cells. A Hungarian assignment is then solved between interactions and labels, with the cost of assigning interaction $k$ a label selecting path $p$ given by $c(k, p) = w_\ell \cdot |p| - w_d \cdot b(p, k)$, where $|p|$ is the path length, $b(p, k)$ is a direction-sharing bonus rewarding paths whose step-direction signature is shared by many other interactions, and $w_\ell, w_d$ are weights. This bonus is the Hungarian analogue of the AOD-reuse term in the SA cost, encouraging globally consistent move directions so more paths can share an AOD.


\subsection{AOD-SLM Transfer Minimization}
\label{loss-aware-compilation}
 

\begin{figure}[t]
    \centering
    \includegraphics[width=0.99\linewidth]{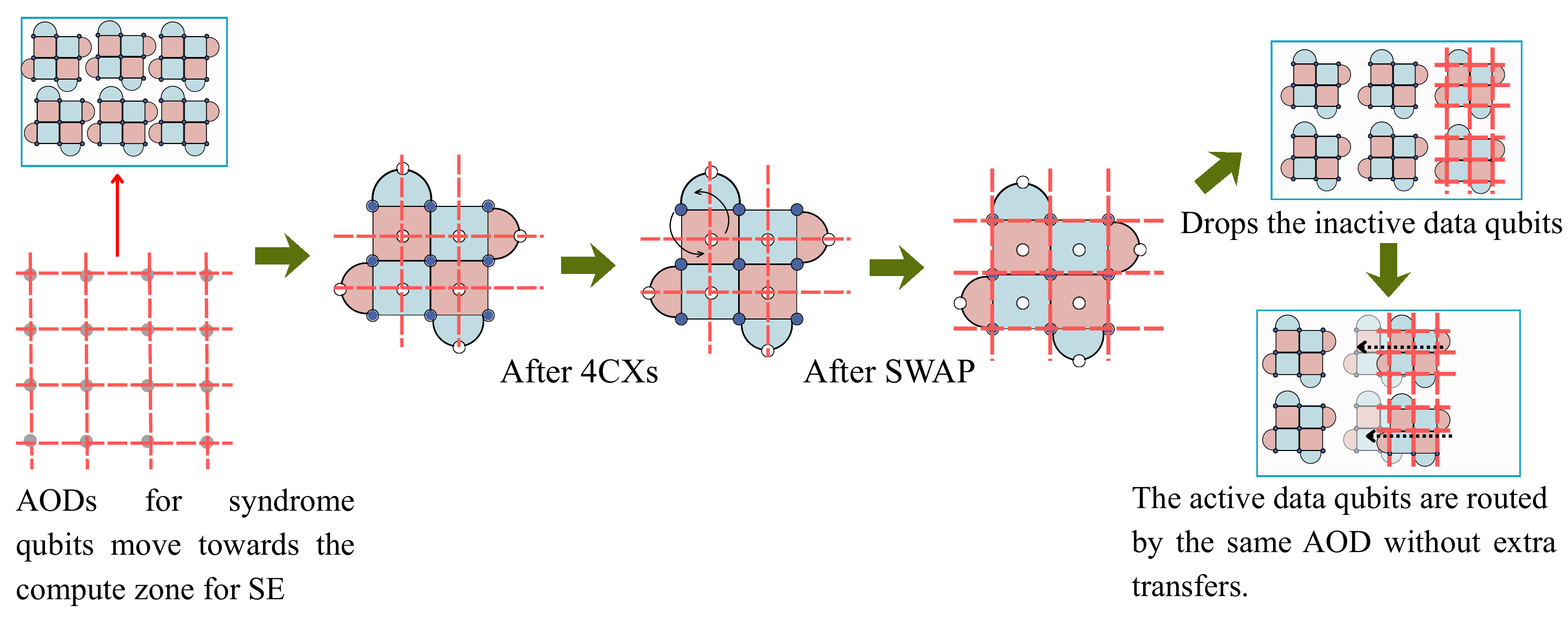}
    \caption{Transfer loss is reduced by selectively dropping inactive qubits after each SE.}
    \label{fig:aod_transfer_loss}
\end{figure}
As discussed in \autoref{sec:transfer_loss} and \autoref{fig:transfer_loss}, transfer loss is a significant contributor to loss accumulation, reaching up to four pickups per transversal gate in the zoned architecture. \textbf{We mitigate this with a transfer-deferred compilation pass where a qubit reused later in the next layer stays in its AOD tweezer instead of returning to its SLM trap.} The pickup count for a contiguous run of 
$k$ operations on the same qubit, therefore, drops from $\sim4k$ to $\leq2k$. The saving scales linearly with the depth of qubit reuse of the circuit structure.

\autoref{fig:aod_transfer_loss} shows a typical AOD-control timeline that realizes this deferral. A set of AODs first moves the ancilla qubits from the storage zone into the compute zone, with the number of AODs dispatched set by how many active gates the routing pass of \autoref{sec:aod_routing} schedules for the next layer. After the stabilizer check, the SWAP SE round physically exchanges the data and ancilla qubits, so each AOD that was holding an ancilla qubit now holds the data qubit it swapped with. The same AOD can therefore carry the data qubits of the \textit{active} logical qubits (those with a transversal gate in the next layer) straight into their next operation, with no intermediate SLM transfer. Only the \textit{inactive} data qubits are released back into their SLM traps, each incurring a single pickup.

 The scheduler (described in \autoref{sec:aod_routing}) determines which operations are active in each layer, while the assigning of each AOD to the operation is determined in this pass. It decides how many atoms must be transferred between traps by solving a linear assignment problem exactly with the Hungarian algorithm. The cost of assigning a set of operations to a device is the number of transfers that choice forces. The transfers are determined by the number of atoms that would have to be moved from the SLM to AOD and atoms already resident in some other AOD device. Under a fixed total AOD budget, this minimizes inter-layer transfers, letting the SWAP-exchanged atoms move straight back into a gate instead of cycling through their SLM traps. The deferral keeps any qubit in transit for a few layers before SE takes place and old atoms are moved to readout. The number of layers in between SE is bounded by the decoder budget $p_{\mathrm{dec}}$. \autoref{sec:overall}

%% file: sections/5-architectural_modelling.tex
\section{Architectural Modelling} \label{sec:evaluation}

\subsection{Loss-correction protocols for surface code} \label{subsec:background-loss-correction}

\subsubsection{Delayed-erasure decoding} \label{subsec:background-delayed-erasure-decoding}
As described in \autoref{subsec:background-loss-errors}, a loss error cancels subsequent gates on the lost qubit, thereby generating correlated errors.
Decoders designed for Pauli errors cannot effectively correct such errors.
Moreover, qubit loss is not detected immediately; SSR detects it only when the qubit is measured.
Hence, we implement a \emph{delayed-erasure decoder}~\cite{baranes2026leveraging} to correct both loss and Pauli errors.

A QEC decoder estimates spacetime error locations from an error syndrome and a detector error model~\cite{derks2025designing}.
A detector error model is a list of error mechanisms, each of which has an associated error probability and a set of detectors that it flips.
We construct the detector error model $\mathrm{DEM}_{\mathrm{final}}=\mathrm{DEM}_{\mathrm{Pauli}}+\sum_i\mathrm{DEM}_{i}$ and pass it to PyMatching~\cite{Higgott2025sparseblossom}.
Here, $\mathrm{DEM}_{\mathrm{Pauli}}$ is the detector error model constructed from Pauli error information without considering loss errors.
Each $\mathrm{DEM}_{i}$ is a detector error model conditioned on loss detection by measurement $i$ and is constructed without considering Pauli errors or loss errors detected by other measurements.
The sum of detector error models is defined as list concatenation.


\begin{figure}[t]
    \centering
    \includegraphics[width=\linewidth]{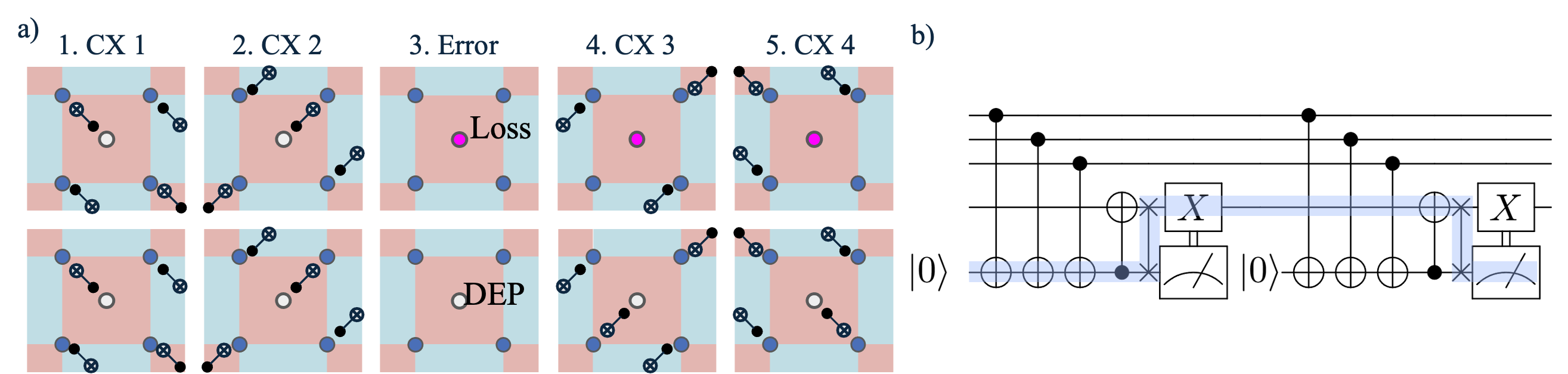}
    \caption{
    a) Effect of an ancilla-qubit loss.
    Blue, white, and pink circles denote data, ancilla, and lost ancilla qubits, respectively.
    The top row shows ancilla loss after the second CX gate; the bottom shows an equivalent circuit with fully depolarizing noise (``DEP'').
    b) Two rounds of swap-based syndrome extraction with a qubit lifecycle highlighted in blue.
    For simplicity, the alternation between the SWAP-pairing patterns is omitted.
    }
    \label{fig:syndrome_extraction_with_loss}
\end{figure}

Given a loss-detecting measurement $i$, the \emph{qubit lifecycle} of $i$ is defined as the set of loss-error locations that can be detected by $i$.
For a specific loss-error location $e$, we construct a detector error model by inserting depolarizing noise that approximates the effects of a loss at $e$.
For example, \autoref{fig:syndrome_extraction_with_loss}(a) shows the case where an ancilla qubit is lost after the second CX gate in the syndrome extraction circuit.
In this case, the effect of the loss error is equivalent to inserting fully depolarizing noise at the loss-error location.
We construct $\mathrm{DEM}_{i}$ by reweighting the noise strengths associated with all loss-error locations in the qubit lifecycle of $i$.


\subsubsection{Swap-based syndrome extraction} \label{subsec:background-swap-syndrome-extraction}
While loss errors on ancilla qubits are detected in every syndrome extraction round and effectively corrected by the delayed-erasure decoder, loss errors on data qubits are more harmful because data qubits are rarely measured and have long qubit lifecycles.
To address this issue, we modify the syndrome extraction circuit to shorten the qubit lifecycles.
Specifically, we use the swap-based syndrome extraction circuit~\cite{baranes2026leveraging} because of its simplicity and modest hardware requirements compared to other schemes.
Combining our architecture with other loss-tolerant decoding schemes, potentially targeting different hardware capabilities, is an interesting direction for future work.
\autoref{fig:syndrome_extraction_with_loss}(b) illustrates the swap-based syndrome extraction circuit.
We exchange an ancilla qubit and a data qubit using an atom-movement-based SWAP, which is immune to gate cancelation caused by loss errors.
Since not all data qubits can be swapped in a single round, we alternate the SWAP-pairing pattern between even and odd SWAP rounds to address this issue.

\subsection{Physical-level modeling}
\begin{scriptsize}
    \begin{table}[t]
      \centering
      \caption{Error model ``NAL'' used in this work.}
      \label{table:parameters}
      \begin{tabular}{|l|l|l|l|l|l|l|l|l|l|}
        \hline
        \textbf{Name} &pickup & transport & meas & 1Q & 2Q & reset \\
        \hline
        \textbf{Loss} & 0.2\% & Eq.~\eqref{eq:heating-loss} & 0.1\% & 0\% & 0.1\% & 0\% \\
        \hline
        \textbf{Pauli} & 0\% & 0\% & 0.1\% & 0\% & 0.05\% & 0.1\% \\
        \hline 
        \textbf{Duration} & 50 \textmu s & Eq.~\eqref{eq:heating} & 500 \textmu s  & 1 \textmu s  & 1 \textmu s & 200 \textmu s  \\
        \hline
        \textbf{Ref} & \cite{Manetsch2024ATA} & \cite{Hwang:25} & \cite{Bluvstein2024} & see text & \cite{Evered2026} & \cite{norcia2023}  \\
        \hline
      \end{tabular}
      \label{tab:noise-parameters}
    \end{table}

\end{scriptsize}

We evaluate atom loss on the output schedule of the compiler. Each atom is exposed to loss at four kinds of events (summarized in \autoref{tab:noise-parameters}, which we call ``NAL'' error model):
\begin{itemize}
    \item \textbf{Pickup:}  $p_\mathrm{pick}=2\times10^{-3}$, charged whenever an atom is lifted from or moved back into the SLM~\cite{Manetsch2024ATA}. 
    \item \textbf{Transport move:} Atom loss during shuttling is a function distance and speed. We model it from the motional heating a trapped atom suffers under finite-time transport.  Following \cite{Hwang:25}, the transport atom-loss probability is evaluated from the cumulative energy distribution. An atom moved a distance $\ell$ in time $t_f$ is heated by motional quanta
    \begin{equation}\label{eq:heating}
    \Delta n = \tfrac{50}{3}m\ell^{2}/(\hbar\omega^{3}t_f^{4}),    
    \end{equation}
    where $m=$ atomic mass, $\omega=$ trap frequency and $\hbar=$ Planck's constant. $\eta = (U - \Delta n\hbar\omega)/k_BT$ is taken as its remaining energy, the difference between the trap depth $U$ and the acquired energy. The corresponding loss probability is given by Eq.(9) of \cite{Hwang:25}
    \begin{equation}\label{eq:heating-loss}
        p(\eta)=(1+\eta+\tfrac12\eta^{2})e^{-\eta}. 
    \end{equation}
    For a Gaussian distribution of atoms about the peak depth $U_0=\tfrac12 m\omega^{2}a^{2}$ where $a$ is the tweezer beam waist, we sample $U$ within a certain standard deviation $\sigma_U $ and report the average loss $p_\mathrm{loss}(\ell,t_f)=\langle p(\eta)\rangle$ over that distribution.
    \item \textbf{Rydberg gate:} $p_\mathrm{gate}=10^{-3}$ loss rate and 0.05\% Pauli error rate per gate~\cite{Evered2026}.
    \item \textbf{Imaging:} $p_\mathrm{img}=10^{-3}$ loss per measurement, with a 0.1\% probability of assigning a wrong measurement~\cite{Bluvstein2024}.
    \item \textbf{Single-qubit gate:} for simplicity, we consider their effect to be negligible compared to other operations. Indeed, near $10^{-4}$ error rate per gate is reported~\cite{Manetsch2024ATA}.
\end{itemize}
\textbf{Total per-cycle loss.} Within a cycle, each active data atom is exposed
independently to pickup, transport, and gate loss, so its survival is the
product of the per-event survivals, and its loss probability is, to a leading order,
\begin{equation}
\begin{aligned}
p_{\mathrm{loss}}
&= 1 - (1-p_{\mathrm{pick}})^{\,n_{\mathrm{pick}}}
(1-p_{\mathrm{mv}})^{\,n_{\mathrm{mv}}}
(1-p_{\mathrm{gate}})^{\,n_{\mathrm{gate}}} \\
&\approx n_{\mathrm{pick}}\,p_{\mathrm{pick}}
+ n_{\mathrm{mv}}\,p_{\mathrm{mv}}
+ n_{\mathrm{gate}}\,p_{\mathrm{gate}},
\end{aligned}
\label{eq:loss_total}
\end{equation}
where $n_{\mathrm{pick}},\,n_{\mathrm{mv}},\,n_{\mathrm{gate}}$ are the pickups, transport moves, and entangling gates the atom undergoes in the cycle. Imaging loss $p_{\mathrm{img}}$ enters the ancilla budget separately.  
%
 
\subsection{Logical error rate modeling}
To model the logical error rate  in the presence of loss, we use Stim~\cite{gidney2021stim} to simulate circuit-level memory experiments on the rotated surface code with the SWAP SE protocol and the delayed-erasure decoder.
For each code distance $d \in \{3,5,7,9\}$, we sample up to $10^{7}$ shots per configuration and record the LER as the fraction of shots with an incorrect logical outcome.
We use the NAL error model for syndrome extraction and separately consider losses between SE rounds to model those occurring during gate execution and qubit shuttling in the computation.

\begin{figure}[t]
    \centering
    \includegraphics[width=\linewidth]{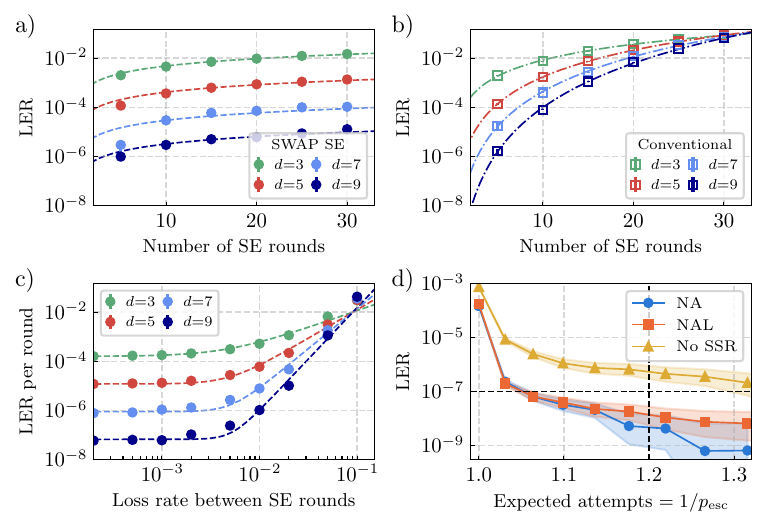}
    \caption{Logical error rate modeling.}
    \label{fig:ler_model}
\end{figure}

Figs.~\ref{fig:ler_model}(a, b) show LERs at an inter-round loss rate of $0.5\%$ for SWAP SE and conventional SE, respectively.
While SWAP SE exhibits a linear increase in LER with the number of SE rounds (dashed lines are the linear fit), conventional SE is unable to correct loss errors and exhibits a rapid increase in LER.
\autoref{fig:ler_model}(c) shows the LER per round for varying inter-round loss rates with the SWAP SE, which we fit with a two-term scaling ansatz
\begin{equation}\label{eq:ler_scaling}
    P_\mathrm{L}(d,p)=A\Big(\tfrac{p}{p_0}\Big)^{\alpha d}+B\gamma^{(d+1)/2},
\end{equation}  
yielding the idling loss threshold $p_0$ and the effective scaling given by the fit parameters $A, B, \gamma, \alpha$.
We adopt this memory-based evaluation as a conservative estimate of the LER
in full logical computation. Following
Ref.~\cite{sunami2025transversalsurfacecodegamepowered}, our instruction set
implements $S$ and $T$ via gate teleportation, which terminates qubit
lifecycles by replacing qubits with freshly prepared logical ancillae. As
Ref.~\cite{baranes2026leveraging} shows, lifecycles in deep logical circuits
are therefore shorter than in memory experiments. Correlated decoding of
transversal Clifford circuits further supports this, since the decoding
problem decomposes into a collection of memory-like decoding
graphs~\cite{cain2025fastcorrelateddecodingtransversal}.

\subsection{Magic-state cultivation in the presence of losses} \label{subsec:msc-loss}

As described in \autoref{subsec:background-ftqc-surface-code}, we use the fold-transversal surface code cultivation~\cite{sahay2025fold} with $\mathrm{f}=3$ to prepare $T\ket{+}$ magic states.
Because loss errors are not considered in the original study,\footnote{Ref.~\cite{jacoby2025magic} analyzes the performance of the injection and cultivation stages of Gidney's magic state cultivation~\cite{gidney2024magic} in the presence of erasure errors. That study, however, does not evaluate the impact of erasure errors on the escape stage, which is where the majority of the physical gates are executed, and therefore the effect of erasure errors is expected to be dominant.} we evaluate the impact of loss errors on the protocol. 
To do so, we simulate the protocol in two parts: the first simulation covers the injection and cultivation stages, and the second covers the escape stage.
We consider two noise models.
The first is the NAL model.
The second, simplified model is the ``NA'' model; this is similar to the ``PM1'' noise model in Ref.~\cite{sahay2025fold}, but ignores single-qubit gate errors.
The two-qubit gate Pauli error rate is set to $10^{-3}$, no loss errors are considered, and each reset/measurement operation is subject to bit-flip or phase-flip noise with strength $10^{-3}$.

The first simulation evaluates the cultivation protocol, targeting $\ket{Y}$ state~\cite{sahay2025fold,gidney2024magic}.
Under the NA model, it yields an LER of $2.1 \times 10^{-8}$ and an acceptance rate of $0.84$.
Under the NAL model, it yields an LER of $8.5 \times 10^{-9}$ and an acceptance rate of $p_\mathrm{cult}=0.78$, where we discard trials with any loss detection or measurement flips~\cite{jacoby2025magic}.
For the escape stage that grows the code distance~\cite{sahay2025fold}, we perform a simulation where we prepare a $\ket{0}$ state encoded in the distance-3 regular surface code in a noise-free manner, convert it to the distance-5 rotated surface code, expand it to distance 7, and then perform a noise-free measurement and postselection based on the complementary gap~\cite{gidney2024magic}.
For the NAL noise model, we utilize SWAP SE in the escape stage and perform loss-aware decoding to calculate the complementary gap.
The results are shown in \autoref{fig:ler_model}(d); The NA and NAL models show similar error suppression as a function of decreased acceptance rate $p_\mathrm{esc}$ (increased ``expected attempts'', $1/p_\mathrm{esc}$) despite the increased total error rates of the NAL model.
Without erasure-aware decoding, the expansion stage with the NAL model results in an order of magnitude larger LER (``No SSR'').
We also performed $X$-basis simulations and observed no appreciable difference in LER; accordingly, we estimate the combined logical $X$- and $Z$-error rate of the escape stage as twice the $Z$-basis LER.
We use these results to conservatively estimate the expected number of cultivation attempts required for a given target LER of the magic state.
For example, we estimate that preparing a magic state with an LER of $10^{-7}$ would be achieved by $1/p_\mathrm{esc} = 1.2$ (dashed lines in \autoref{fig:ler_model}(d)), resulting in a total cultivation trial count of $1/(p_\mathrm{esc}p_\mathrm{cult}) = 1.53$.
\footnote{For the cultivation stage, we used a $\ket{Y}$ state cultivation protocol as a proxy for magic state cultivation, as in Ref.~\cite{sahay2025fold} where the authors reported nearly identical LER for both $\ket{Y}$ and $\ket{T}$ state cultivation at f=3. Since careful assessment is required to ensure their conclusion carries over to other error models, we used conservative LER and postselection cost estimates here. 
Furthermore, we simulated the cultivation and escape stages separately and combined the results conservatively, instead of a combined 'hand-off' simulation of Ref.~\cite{sahay2025fold}.}


\subsection{Overall performance estimation}
\label{sec:overall}
 
The loss a program incurs is read directly from its compiled schedule using \autoref{eq:loss_total}; 
averaging the per-atom loss over the atoms active in a layer, and then over layers (number of SE rounds), yields a single per-round loss rate  $p_\mathrm{rnd}(d)$ for qubits. 
Shuttling lengths and operation counts affect the loss and are entirely obtained from the compilation:
shorter, more local routing produces shorter shuttling and smaller $p_\mathrm{rnd}$. 
For the SWAP SE protocol which interchanges the role of ancilla and data qubits, we can approximate $p_\mathrm{rnd}(d)$ as the mean of per-layer ancilla and data qubit loss probability. 


The fitting of the decoder simulation with \autoref{eq:ler_scaling} provides the maximum per-round loss that is tolerated. The compiled loss rate is equated to the decoder's loss budget (via inter-round loss probability $p_\mathrm{rnd}(d)$) to establish feasibility. 
At each candidate distance, the compiled $p_\mathrm{rnd}(d)$ is compared to the decoder's per-round loss $p_\mathrm{dec}(d)$ the surface code can absorb at the target LER, and the smallest $d$ satisfying $ \min_d(p_\mathrm{rnd}(d)\le p_\mathrm{dec}(d))$
is the minimum feasible code distance $d_{min}$. Since $p_\mathrm{rnd}$ grows with $d_{min}$ (transports lengthen as $\ell\propto d$), the matching fixes $d_{min}$ for each benchmark, and a compiler with shorter transports meets the budget at a smaller $d_{min}$.

For non-Clifford gates, we first need to set an overall error budget, which is determined primarily by the application requirements.
Once the total error budget allocated to $T$ gates is determined, the target LER for each $\ket{T}$ magic state can be determined by dividing the allocated budget by the $T$ count.
As described in \autoref{subsec:msc-loss}, this target LER determines the required number of cultivation attempts, which is then used to calculate the number of cultivation modules in the system.

%% file: sections/6-experimental_setup.tex
\section{Experimental Setup}
\subsection{Benchmarks}
We use applications of different scales and routing requirements~\cite{obenland_2026_18154991,li2022qasmbench}. We consider first-order Trotterized time-evolution circuits for lattice Hamiltonians such as the transverse Ising, Heisenberg, and Fermi-Hubbard models, at scales ranging from 4 to 100 qubits and across different numbers of Trotter steps \(N_{tr}\). We also study qubitized ground-state energy estimation (GSEE) circuits, which use qubitized phase estimation over a PauliLCU block encoding for smaller instances\cite{obenland_2026_18154991}. We include a set of standard structured circuits, such as the adder and W-state preparation circuits\cite{li2022qasmbench}. 
\begin{table}[!t]
\caption{Benchmarks}
\label{table:benchmarks}
\centering 

\renewcommand{\arraystretch}{1.00}
\setlength{\tabcolsep}{2pt}
\begin{tabular}{@{}l r r r r@{}}
\toprule
\textbf{Program} & \textbf{T ops} & \textbf{CX ops}  & \textbf{Total Ops} & \textbf{Qubits} \\
\midrule
\multicolumn{4}{@{}l}{\textit{Trotterized dynamics}} \\
\midrule
\quad Ising ($N_{\mathrm{steps}}{=}198$)          & $91{,}036$ & $1,584$ & $252{,}153$   & 4   \\
\quad Ising ($N_{\mathrm{steps}}{=}50$)          & $268,800$ & $5,200 $& $697,200$   & 32  \\
\quad Ising ($N_{\mathrm{steps}}{=}2$)          & $18{,}304$ &$448$ & $48{,}576$   & 64   \\
\quad Ising ($N_{\mathrm{steps}}{=}1$)            & $14{,}000$ &$360$ & $36{,}960$    & 100 \\
\quad Heisenberg ($N_{\mathrm{steps}}{=}290$)     & $350{,}260$ &$3,480$&$928{,}160$   & 4   \\
\quad Heisenberg ($N_{\mathrm{steps}}{=}50$)     & $738,800$ &$7,800$&  $1,931,000$   & 32   \\
\quad Heisenberg ($N_{\mathrm{steps}}{=}2$)     & $50{,}944$ & $672$&$135{,}872$   & 64   \\
\quad Heisenberg ($N_{\mathrm{steps}}{=}1$)       & $37{,}400$ &$540$ & $102{,}320$   & 100 \\
\quad Fermi-Hubbard ($N_{\mathrm{steps}}{=}588$) & $423{,}564$ & $4,704$&$1{,}150{,}036$ & 4 \\
\quad Fermi-Hubbard ($N_{\mathrm{steps}}{=}5$) & $40,800$ & $4,716$ &$109,680$ & 32 \\
\quad Fermi-Hubbard ($N_{\mathrm{steps}}{=}2$) & $32{,}384$ & $4,271$&$ 89{,}280$ & 64 \\
\quad Fermi-Hubbard ($N_{\mathrm{steps}}{=}1$)   & $25{,}180$ &$4,170$ & $72{,}220$    & 100 \\
\addlinespace[2pt]
\midrule
\multicolumn{4}{@{}l}{\textit{Ground-state energy estimation (GSEE)}} \\
\midrule
\quad $H_{2}$ molecule        & $13{,}648$&$2,558$ & $56{,}502$ & 19 \\
\quad Fermi-Hubbard ($N{=}8$) & $8{,}288$ &$4,390$ & $27{,}080$ & 25 \\
\quad Heisenberg ($N{=}4$)     & $8{,}224$ &$4,262$ & $59{,}562$ & 21 \\
\addlinespace[2pt]
\midrule
\multicolumn{4}{@{}l}{\textit{State preparation}} \\
\midrule
\quad W-state & $2{,}646$ &$52$& $7{,}145$ & 27 \\
\quad Adder & $96$ &$195$&$616$ & 27 \\

\bottomrule
\end{tabular}
\end{table}
 
 \subsection{Baseline}
Our baseline is the zoned architecture compiled with its native
zoned-architecture compiler\cite{10946298}. We compare it against our
loss-aware compilation(~\autoref{sec:loss_aware_architecture}). Each architecture is
paired with the compiler designed for it, and we do not mix a compiler
across architectures. Our routing passes target the AOD-driven
movement structure of our layout and are not defined on the zoned
layout.

%% file: sections/7-evaluation.tex
\section{Results}

\subsection{Transfer-Deferred Loss Comparison }
 
\begin{figure}[t]
    \centering
    \includegraphics[width=\linewidth]{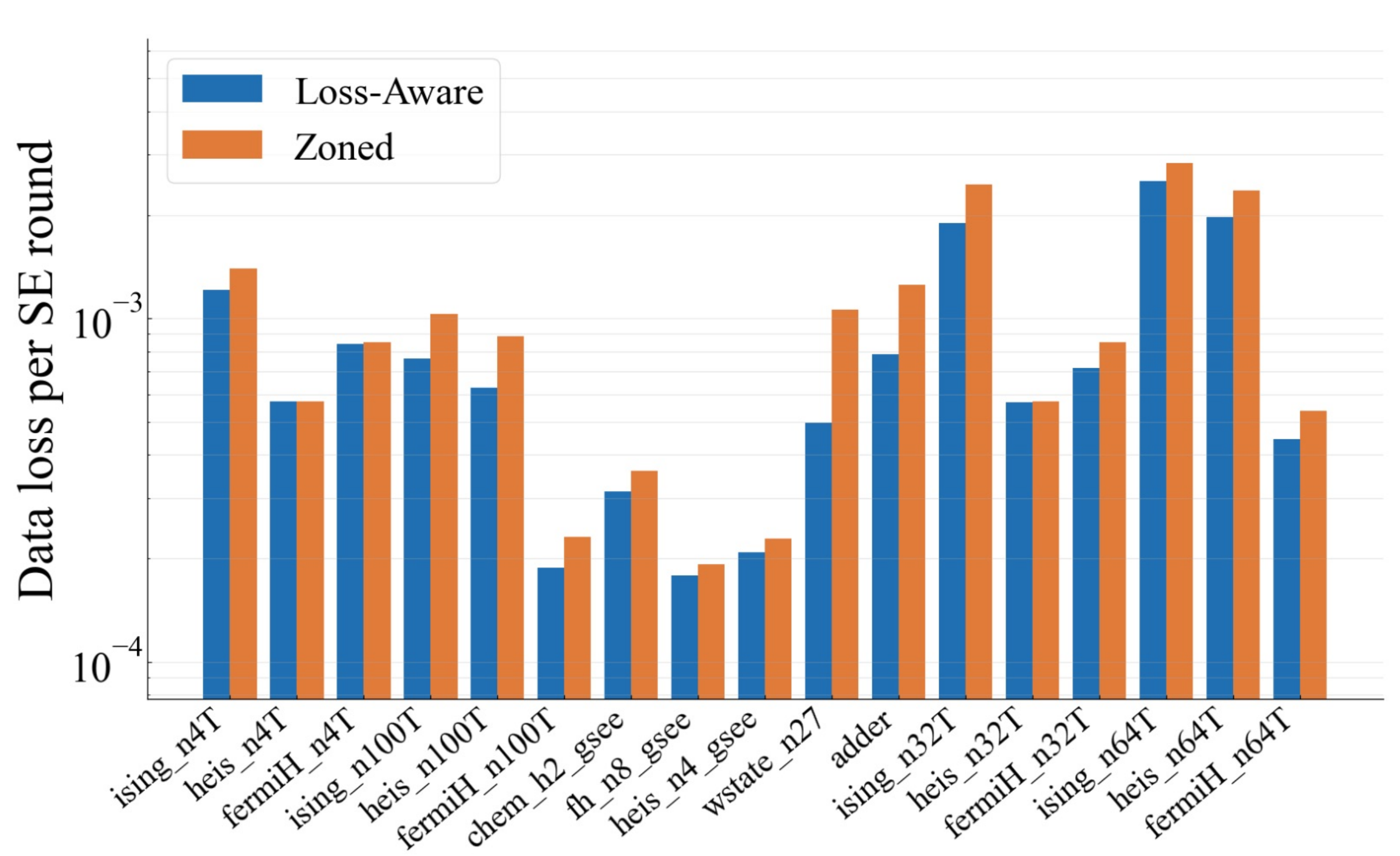}
    \caption{Transfer-deferred loss comparison in zoned and loss-aware layout (lower is better) with a mean reduction of 1.45×; highest gains are on high-reuse circuits such as wstate and adder.}
    \label{fig:loss_aware}
\end{figure}

Fig.~\ref{fig:loss_aware} compares the per-SE-round loss accumulated under the loss-aware approach against zoned-architecture loss accumulation.
Per-SE-round loss is the expected number of atoms lost per logical qubit over one full round of syndrome extraction. Selective transfers between traps lower the loss on every benchmark, with the reduction scaling with a circuit's temporal locality. The gain is largest for the shallow, high-reuse circuits, reaching 2.15× (53$\%$ lower) for wstate and 1.60× (37$\%$) for adder. It reaches 1.13–1.41× on the transport-heavy 2D lattices (ising/heis n64–n100).\textbf{ Across all 17 benchmarks the mean reduction in the data channel is 1.25×, confirming that keeping atoms resident in the AOD removes a substantial fraction of the per-round pickup loss.}


\begin{figure}[!t]
  
    \centering
    \includegraphics[width=\linewidth]{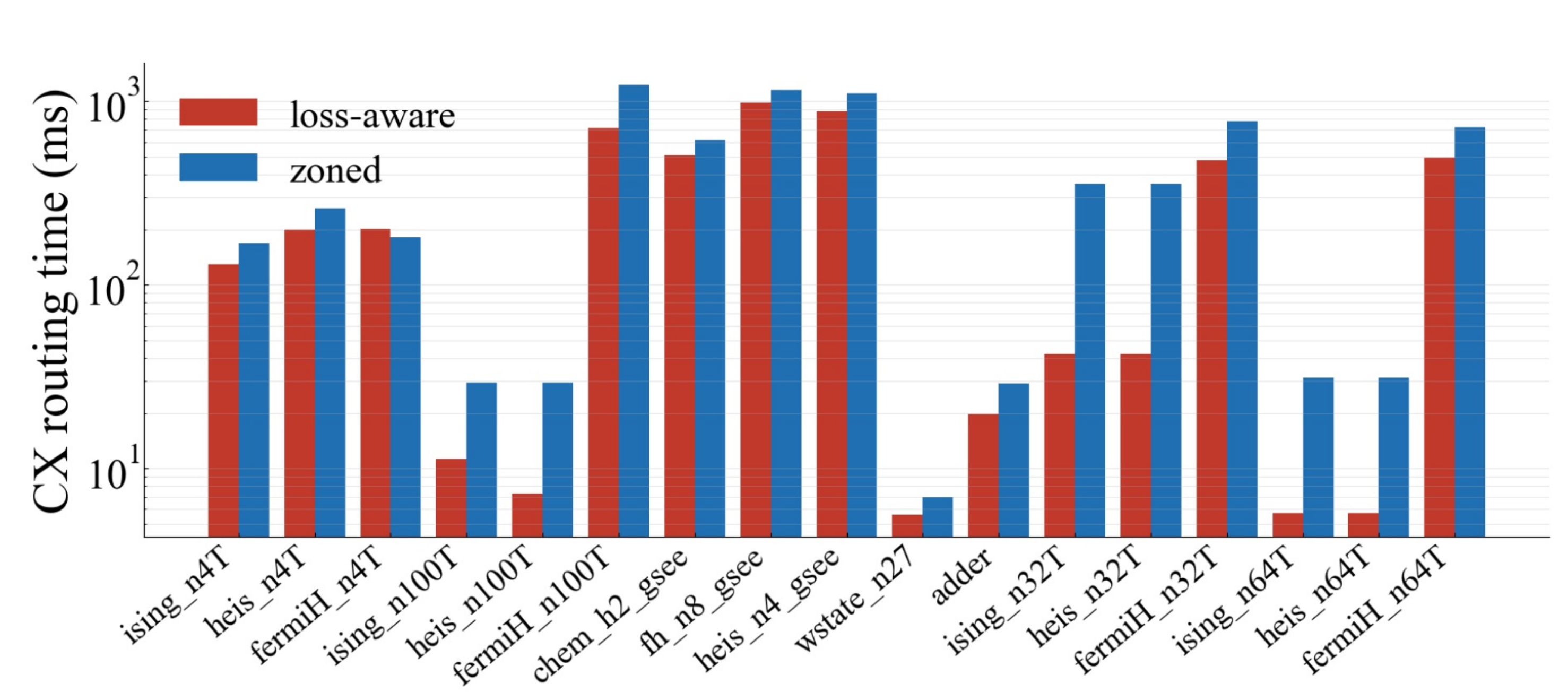}
    \caption{Routing time comparison in zoned and loss-aware layout (log scale, lower is better). The highest benefits of parallelism are seen in the 32- and 64-qubit lattice circuits, with reductions of $88\%$ and $82\%$ in CX-routing time. Number of AODs for computation = 4.}
    \label{fig:cx_routing}

  \label{fig:loss_aware_optimization_results}
\end{figure}
 
\subsection{Loss-Aware vs. Zoned Architecture}
 
We benchmark the AOD-routing pass (\autoref{sec:aod_routing}) on our
loss-aware layout against ZAC~\cite{10946298}, a zoned rearrangement compiler in which qubits reside in a storage zone. For each entangling layer, the required pairs are rearranged by the AOD into an adjacent
entangling zone, subjected to a Rydberg CX, and returned. ZAC is not fault-tolerant and generates physical atom rearrangements; we scale it to the fault-tolerant setting by assuming comparable site spacings and treating
each rearrangement as a token for logical-patch movement. Since ZAC accounts only for CX-transport rearrangements while our compiler also routes T-states, we evaluate only the CX layers under ASAP scheduling, assessing both on the same metric.

Figure~\ref{fig:cx_routing} reports CX-routing time for our benchmarks. Our parallel routing pass attains the lowest transport time on 10 of 11 benchmarks. The margin is largest on the Trotter lattices, where parallel layering exploits the many spatially disjoint two-qubit gates. \textbf{On the 32-qubit Ising circuit it routes in $42.0$~ms against ZAC's $353.4$~ms, an $8.4\times$ reduction. CX routing time falls by $2.9\times$ on average in loss-aware as opposed to zoned layouts.}  One exception is $N=4$ Fermi-Hubbard circuit, which has little disjoint parallelism to exploit. 

\autoref{tab:ler} compares loss-aware vs. zoned architecture logical error rates at a fixed code distance. For each circuit we determine the distance $d$ needed to reach the target LER, then evaluate both compilers on their respective architectures at that $d$, reporting $\rho = \mathrm{LER}_{\mathrm{zoned}}/\mathrm{LER}_{\mathrm{loss\text{-}aware}}$ (higher is better). On near-Clifford and small circuits the two are comparable ($\rho\approx1$-$2$ ), since routing contributes little loss. \textbf{The advantage grows sharply on the 2D-lattice circuits which are important for quantum advantage demonstrations. ZAC's longer movements accumulate transport loss: for Heisenberg circuit, $\rho$ rises from 10.1 (n32) to 13.4 (n64) to 212.4 (n100).} This reflects that the loss-aware layout's shorter, transfer-minimized movements keep per-round loss below the correction threshold.

     
\begin{table}[t]
\caption{The LER advantage grows with larger circuits}
\label{tab:ler}
\centering
\small
\setlength{\tabcolsep}{5pt}
\begin{tabular}{@{}lccc r@{}}
\toprule
Circuit & $d$ & Loss-aware & Zoned & $\rho$ \\
\midrule
chem\_h2\_gsee   & 11 & $8.25{\times}10^{-9}$  & $9.99{\times}10^{-9}$  &   1.2 \\
adder            &  7 & $1.51{\times}10^{-6}$  & $1.89{\times}10^{-6}$  &   1.3 \\
heis\_n4\_gsee   & 11 & $7.73{\times}10^{-9}$  & $1.25{\times}10^{-8}$  &   1.6 \\
fh\_n8\_gsee     & 11 & $8.87{\times}10^{-9}$  & $1.45{\times}10^{-8}$  &   1.6 \\
fermiH\_n32T     & 11 & $1.16{\times}10^{-8}$  & $3.39{\times}10^{-8}$  &   2.9 \\
fermiH\_n64T     & 11 & $1.18{\times}10^{-8}$  & $7.43{\times}10^{-8}$  &   6.3 \\
ising\_n32T      & 13 & $4.51{\times}10^{-10}$ & $4.06{\times}10^{-9}$  &   9.0 \\
heis\_n32T       & 13 & $4.51{\times}10^{-10}$ & $4.55{\times}10^{-9}$  &  10.1 \\
ising\_n64T      & 11 & $6.31{\times}10^{-9}$  & $8.45{\times}10^{-8}$  &  13.4 \\
heis\_n64T       & 11 & $6.31{\times}10^{-9}$  & $8.45{\times}10^{-8}$  &  13.4 \\
ising\_n100T     & 11 & $2.68{\times}10^{-8}$  & $1.39{\times}10^{-6}$  &  52.0 \\
fermiH\_n100T    & 11 & $1.43{\times}10^{-8}$  & $1.15{\times}10^{-6}$  &  80.6 \\
heis\_n100T      & 11 & $6.56{\times}10^{-9}$  & $1.39{\times}10^{-6}$  & 212.4 \\
\bottomrule
\end{tabular}
\end{table}

\subsection{Informing Neutral Atom Hardware Targets}
We derive concrete device implementation targets for the motional-excitation budget, AOD count, and atom reloading rate. While these parameters are available from current demonstrations, prior works do not provide guidance on how future systems must tune these knobs.
 \begin{figure}[t]
    \centering
    \includegraphics[width=1.0\linewidth]{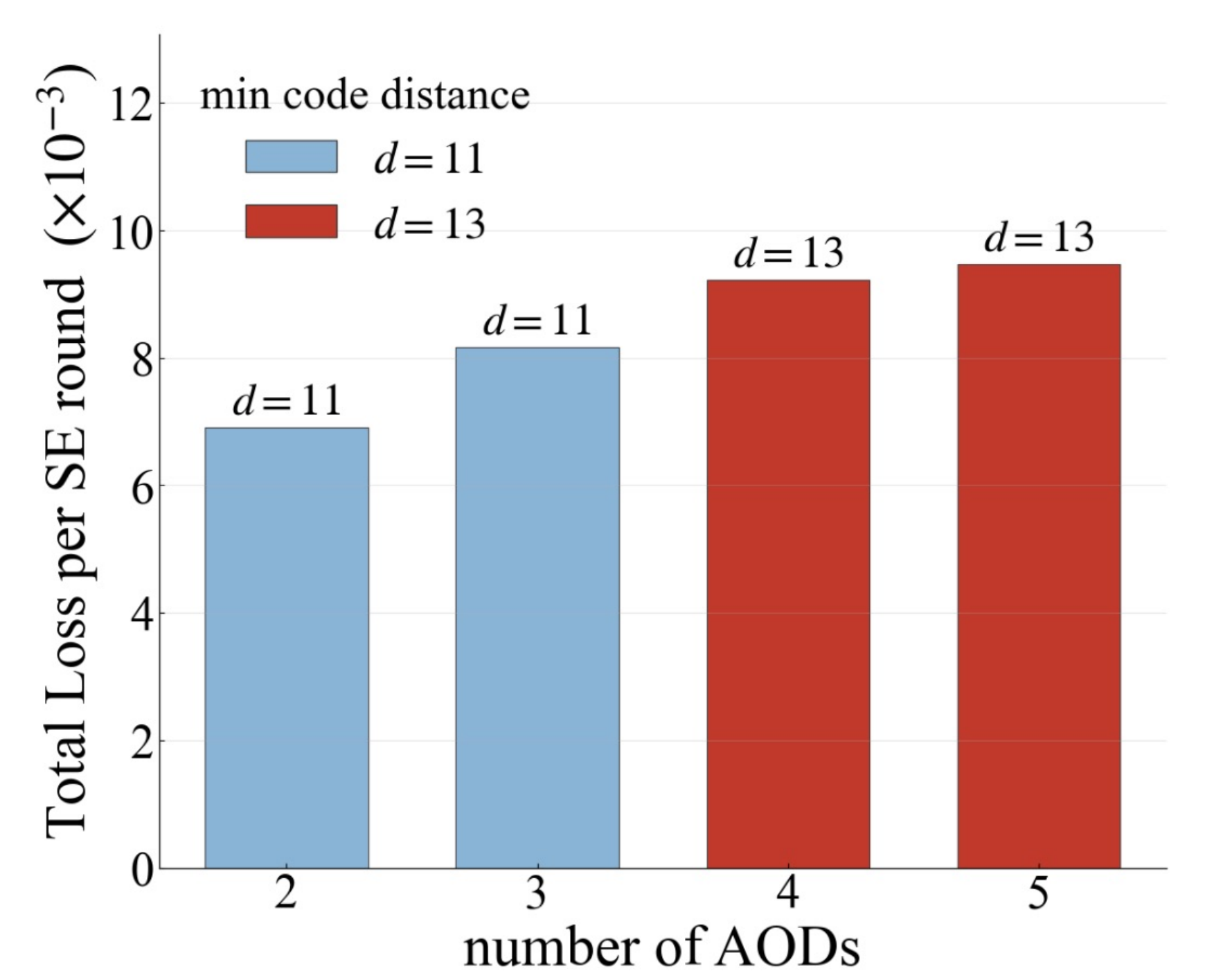}
     
    \caption{Ising Trotterized ($N=64$) circuit: Loss per SE round v/s number of AODs (at $\Delta{n}=2$) }
    \label{fig:aod_sweep}
\end{figure}
\subsubsection{\texorpdfstring{$\Delta{n}$}{delta\_n} Dependence}
\begin{figure}[t]
    \centering
     \includegraphics[width=1.0\linewidth]{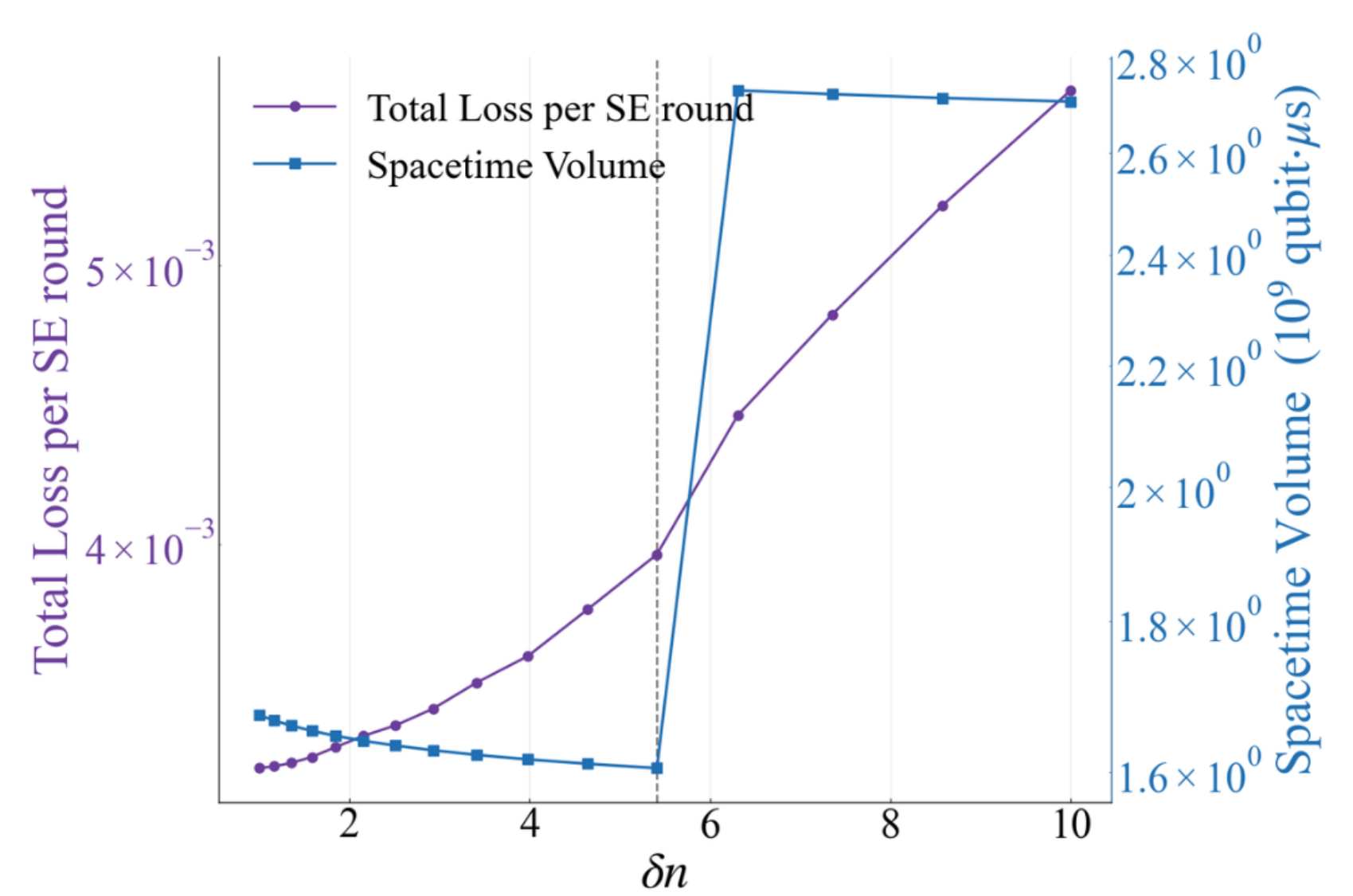}
     \caption{Ising Trotterized ($N=100$) circuit: Loss per SE round (and spacetime volume) v/s motional excitation budget ($\Delta{n}$), Minimum at $\Delta n\sim5$ ($d=11$).}
     \label{fig:deltan}
\end{figure}

To quantify how fault-tolerant resource cost depends on transport speed, we
sweep the motional-excitation budget $\Delta n$ (\autoref{eq:heating}). At
each value of $\Delta n$, we take the smallest code distance whose per-SE-round  loss ($p_{rnd}$)
remains within the code's tolerable budget, obtained by comparing accumulated
inter-round loss against the decoder-simulation fit ($p_{dec}$)
(\autoref{eq:ler_scaling}; see also \autoref{sec:overall}). Since faster moves heat the atoms more, $d_{\min}(\Delta n)$ is a non-decreasing step function of speed (\autoref{fig:deltan}). At small $\Delta n$, the transport loss is relatively less leading to a small code distance. As $\Delta n$ grows, the loss crosses the budgets of successively larger codes.

Compiling each benchmark at $d_{\min}(\Delta n)$ and measuring its execution time yields the space-time volume
 $ V(\Delta n) = d_{\min}(\Delta n)^{2}\, n_{\mathrm{logical}}\,
  t_{\mathrm{exec}}(\Delta n).$ Larger $\Delta n$ lowers $t_{\mathrm{exec}}$ but forces $d_{\min}$ upward in
discrete steps, so $V$ decreases slowly while the distance holds and jumps sharply where it changes. The two effects balance at an interior optimum, here at $\Delta n \sim 5$ (\autoref{fig:aod_sweep}). The sweep therefore delineates a feasible window of $\Delta n$ and the qubit cost of transport at any given speed. Though swept on the Fermi-Hubbard Trotterized circuit, the mechanism holds suite-wide

\subsubsection{Number of AODs}
Figure~\ref{fig:aod_sweep} reports the atom-loss probability per-SE-round for the 100-qubit Ising circuit as the number of AOD arrays is swept from 2 to 5. Adding AODs lets the router transport more atoms per layer, packing more gates into each round and compressing the schedule but concentrating the same total loss into denser rounds, so the per-round loss rises monotonically. This propagates to the required code distance: the loss stays within a $d=11$ code for 3 AODs but crosses its budget at 4, forcing $d=13$, where it then remains through 5 AODs. Additional AOD parallelism shortens the schedule at the cost of a larger code distance.  

 \subsubsection{Reloading Rates}
 Reloading replenishes atoms mid-circuit from a continuously fed cold-atom reservoir  by moving atoms from the reservoir to storage during computation \cite{Li2025FastCA}. 
Table \ref{tab:loading} reports the reloading rate our compiled circuits demand across the benchmark suite. The required rates span roughly three orders of magnitude, from $2.24\times10^{4}$ atoms/s for the smaller circuits to $10^{6}-10^{7}$ for Trotterized lattice circuit benchmarks. Recent demonstrations of continuous mid-circuit atom replacement \cite{Li2025FastCA} achieve extraction of the order of $10^{4}$, with a projected reservoir ceiling of $\sim10^{6}$ atoms/s.
 \textbf{For FTQC implementation, reloading throughput is the parameter that must improve by orders of magnitude with respect to the predicted numbers in the literature.} 


\begin{table}[t]
\caption{Atom Reloading Rate Required per Benchmark ($k=4$)}
\label{tab:loading}
\centering
\small
\setlength{\tabcolsep}{6pt}
\begin{tabular}{@{}lr@{\hspace{2em}}lr@{}}
\toprule
Circuit & Rate (atoms/s) & Circuit & Rate (atoms/s) \\
\midrule
ising\_n4T    & $1.10{\times}10^{6}$ & fermiH\_n4T    & $4.47{\times}10^{6}$ \\
ising\_n32T   & $1.53{\times}10^{7}$ & fermiH\_n32T   & $5.06{\times}10^{5}$ \\
ising\_n64T   & $7.18{\times}10^{6}$ & fermiH\_n64T   & $4.48{\times}10^{5}$ \\
ising\_n100T  & $3.55{\times}10^{6}$ & fermiH\_n100T  & $3.60{\times}10^{5}$ \\
heis\_n4T     & $1.90{\times}10^{6}$ & chem\_h2\_gsee & $1.32{\times}10^{5}$ \\
heis\_n32T    & $5.04{\times}10^{6}$ & fh\_n8\_gsee   & $6.19{\times}10^{4}$ \\
heis\_n64T    & $3.35{\times}10^{6}$ & heis\_n100T   & $2.78{\times}10^{6}$\\
 adder          & $2.24{\times}10^{4}$ & wstate\_n27    & $7.65{\times}10^{5}$ \\
             
\bottomrule
\end{tabular}
\end{table}

%% file: sections/8-discussion.tex
\section{Conclusions}

In this paper, we have addressed the gap between system-level FTQC
architecture studies, which largely ignore qubit loss, and loss-tolerant
protocols, which have not been evaluated end-to-end on a compiled program.
A crucial aspect of our work is treating loss not as a predetermined error
parameter but as a budget spent across a whole program so that layout,
compilation, and decoding can be co-designed to control it. 
We achieve this
by moving away from the zoned approach, where long-distance shuttling and
repeated SLM-AOD handoffs dominate the loss budget, and instead executing
all logical operations in a single compute zone using spatially selective
gates, coupled with AOD-configuration-aware routing that maximizes gate
parallelism within the RF tone budget and a transfer-deferred schedule that
keeps reused atoms resident in the AOD. 
This reduces the accumulated loss per syndrome-extraction round by $1.25\times$ on average (up to $2.15\times$), and  CX routing time by up to $8.5\times$ compared to the baseline architecture. 
Combined with complete logical-level evaluation with a delayed-erasure decoder as well as our loss-aware magic state cultivation protocol, the compiler improvement reduces the logical error rate by over two orders of magnitude against the baseline, with the advantage growing with system size.
Beyond these gains, the framework converts application requirements into concrete device targets, such as continuous reloading rates, AOD counts, and shuttling trajectories. 
We expect our compilation strategy to serve as a crucial toolkit for system architects as neutral-atom hardware scales.


%% file: refs.bib
@article{lee2021even,
	title = {Even {More} {Efficient} {Quantum} {Computations} of {Chemistry} {Through} {Tensor} {Hypercontraction}},
	volume = {2},
	url = {https://link.aps.org/doi/10.1103/PRXQuantum.2.030305},
	doi = {10.1103/PRXQuantum.2.030305},
	number = {3},
	journal = {PRX Quantum},
	author = {Lee, Joonho and Berry, Dominic W. and Gidney, Craig and Huggins, William J. and McClean, Jarrod R. and Wiebe, Nathan and Babbush, Ryan},
	year = {2021},
	pages = {030305},
}

@article{madjarov2020high,
  title={High-fidelity entanglement and detection of alkaline-earth Rydberg atoms},
  author={Madjarov, Ivaylo S and Covey, Jacob P and Shaw, Adam L and Choi, Joonhee and Kale, Anant and Cooper, Alexandre and Pichler, Hannes and Schkolnik, Vladimir and Williams, Jason R and Endres, Manuel},
  journal={Nature Physics},
  volume={16},
  number={8},
  pages={857--861},
  year={2020},
  publisher={Nature Publishing Group UK London}
}

@article{Falconi2025,
  title = {Microsecond-Scale High-Survival and Number-Resolved Detection of Ytterbium Atom Arrays},
  author = {Muzi Falconi, A. and Panza, R. and Sbernardori, S. and Forti, R. and Klemt, R. and Abdel Karim, O. and Marinelli, M. and Scazza, F.},
  journal = {Phys. Rev. Lett.},
  volume = {135},
  issue = {20},
  pages = {203402},
  numpages = {9},
  year = {2025},
  month = {Nov},
  publisher = {American Physical Society},
  doi = {10.1103/n3bg-7yw7},
  url = {https://link.aps.org/doi/10.1103/n3bg-7yw7}
}

@article{Norcia2023,
  title = {Midcircuit Qubit Measurement and Rearrangement in a $^{171}\mathrm{Yb}$ Atomic Array},
  author = {Norcia, M. A. and Cairncross, W. B. and Barnes, K. and Battaglino, P. and Brown, A. and Brown, M. O. and Cassella, K. and Chen, C.-A. and Coxe, R. and Crow, D. and Epstein, J. and Griger, C. and Jones, A. M. W. and Kim, H. and Kindem, J. M. and King, J. and Kondov, S. S. and Kotru, K. and Lauigan, J. and Li, M. and Lu, M. and Megidish, E. and Marjanovic, J. and McDonald, M. and Mittiga, T. and Muniz, J. A. and Narayanaswami, S. and Nishiguchi, C. and Notermans, R. and Paule, T. and Pawlak, K. A. and Peng, L. S. and Ryou, A. and Smull, A. and Stack, D. and Stone, M. and Sucich, A. and Urbanek, M. and van de Veerdonk, R. J. M. and Vendeiro, Z. and Wilkason, T. and Wu, T.-Y. and Xie, X. and Zhang, X. and Bloom, B. J.},
  journal = {Phys. Rev. X},
  volume = {13},
  issue = {4},
  pages = {041034},
  numpages = {12},
  year = {2023},
  month = {Nov},
  publisher = {American Physical Society},
  doi = {10.1103/PhysRevX.13.041034},
  url = {https://link.aps.org/doi/10.1103/PhysRevX.13.041034}
}

@article{Henriet2020quantumcomputing,
  doi = {10.22331/q-2020-09-21-327},
  url = {https://doi.org/10.22331/q-2020-09-21-327},
  title = {Quantum computing with neutral atoms},
  author = {Henriet, Lo{\"{i}}c and Beguin, Lucas and Signoles, Adrien and Lahaye, Thierry and Browaeys, Antoine and Reymond, Georges-Olivier and Jurczak, Christophe},
  journal = {{Quantum}},
  issn = {2521-327X},
  publisher = {{Verein zur F{\"{o}}rderung des Open Access Publizierens in den Quantenwissenschaften}},
  volume = {4},
  pages = {327},
  month = sep,
  year = {2020}
}

@article{Sunami2025,
  title = {Scalable Networking of Neutral-Atom Qubits: Nanofiber-Based Approach for Multiprocessor Fault-Tolerant Quantum Computers},
  author = {Sunami, Shinichi and Tamiya, Shiro and Inoue, Ryotaro and Yamasaki, Hayata and Goban, Akihisa},
  journal = {PRX Quantum},
  volume = {6},
  issue = {1},
  pages = {010101},
  numpages = {22},
  year = {2025},
  month = {Feb},
  publisher = {American Physical Society},
  doi = {10.1103/PRXQuantum.6.010101},
  url = {https://link.aps.org/doi/10.1103/PRXQuantum.6.010101}
}

@article{kim2022fault-tolerant,
	title = {Fault-tolerant resource estimate for quantum chemical simulations: {Case} study on {Li}-ion battery electrolyte molecules},
	volume = {4},
	shorttitle = {Fault-tolerant resource estimate for quantum chemical simulations},
	url = {https://link.aps.org/doi/10.1103/PhysRevResearch.4.023019},
	doi = {10.1103/PhysRevResearch.4.023019},
	number = {2},
	journal = {Physical Review Research},
	author = {Kim, Isaac H. and Liu, Ye-Hua and Pallister, Sam and Pol, William and Roberts, Sam and Lee, Eunseok},
	year = {2022},
	pages = {023019},
}

@article{gidney2021how,
	title = {How to factor 2048 bit {RSA} integers in 8 hours using 20 million noisy qubits},
	volume = {5},
	url = {https://quantum-journal.org/papers/q-2021-04-15-433/},
	doi = {10.22331/q-2021-04-15-433},
	journal = {Quantum},
	author = {Gidney, Craig and Ekerå, Martin},
	year = {2021},
	pages = {433},
}

@article{Chen2026,
  title = {Transversal Logical Clifford Gates on the Rotated Surface Code with Reconfigurable Neutral Atom Arrays},
  author = {Chen, Zi-Han and Chen, Ming-Cheng and Lu, Chao-Yang and Pan, Jian-Wei},
  journal = {Phys. Rev. Lett.},
  volume = {136},
  issue = {13},
  pages = {130601},
  numpages = {8},
  year = {2026},
  month = {Mar},
  publisher = {American Physical Society},
  doi = {10.1103/m7tq-9v3g},
  url = {https://link.aps.org/doi/10.1103/m7tq-9v3g}
}

@misc{cain2025fastcorrelateddecodingtransversal,
      title={Fast correlated decoding of transversal logical algorithms}, 
      author={Madelyn Cain and Dolev Bluvstein and Chen Zhao and Shouzhen Gu and Nishad Maskara and Marcin Kalinowski and Alexandra A. Geim and Aleksander Kubica and Mikhail D. Lukin and Hengyun Zhou},
      year={2025},
      eprint={2505.13587},
      archivePrefix={arXiv},
      primaryClass={quant-ph},
      url={https://arxiv.org/abs/2505.13587}, 
}

@article{zhou2025low,
  title={Low-overhead transversal fault tolerance for universal quantum computation},
  author={Zhou, Hengyun and Zhao, Chen and Cain, Madelyn and Bluvstein, Dolev and Maskara, Nishad and Duckering, Casey and Hu, Hong-Ye and Wang, Sheng-Tao and Kubica, Aleksander and Lukin, Mikhail D},
  journal={Nature},
  volume={646},
  number={8084},
  pages={303--308},
  year={2025},
  publisher={Nature Publishing Group UK London}
}

@article{shor1997polynomial-time,
	title = {Polynomial-{Time} {Algorithms} for {Prime} {Factorization} and {Discrete} {Logarithms} on a {Quantum} {Computer}},
	volume = {26},
	issn = {0097-5397},
	url = {https://epubs.siam.org/doi/10.1137/S0097539795293172},
	doi = {10.1137/S0097539795293172},
	number = {5},
	journal = {SIAM Journal on Computing},
	author = {Shor, Peter W.},
	year = {1997},
	pages = {1484--1509},
}

@article{Saffman2016,
author = {Saffman, M},
year = {2016},
month = {10},
pages = {},
title = {Quantum computing with atomic qubits and Rydberg interactions: Progress and challenges},
volume = {49},
journal = {Journal of Physics B: Atomic, Molecular and Optical Physics},
doi = {10.1088/0953-4075/49/20/202001}
}

@misc{sunami2025transversalsurfacecodegamepowered,
      title={Transversal Surface-Code Game Powered by Neutral Atoms}, 
      author={Shinichi Sunami and Akihisa Goban and Hayata Yamasaki},
      year={2025},
      eprint={2506.18979},
      archivePrefix={arXiv},
      primaryClass={quant-ph},
      url={https://arxiv.org/abs/2506.18979}, 
}

@article{Bluvstein:2025ped,
    author = "Bluvstein, Dolev and others",
    title = "{A fault-tolerant neutral-atom architecture for universal quantum computation}",
    eprint = "2506.20661",
    archivePrefix = "arXiv",
    primaryClass = "quant-ph",
    doi = "10.1038/s41586-025-09848-5",
    journal = "Nature",
    volume = "649",
    number = "8095",
    pages = "39--46",
    year = "2026",
    note = "[Erratum: Nature 650, E3 (2026)]"
}

@article{endres2016,
  title={Atom-by-atom assembly of defect-free one-dimensional cold atom 
         arrays},
  author={Endres, M. and Bernien, H. and Keesling, A. and Levine, H. and 
          Anschuetz, E. R. and Krajenbrink, A. and Senko, C. and Vuleti{\'c}, 
          V. and Greiner, M. and Lukin, M. D.},
  journal={Science},
  volume={354},
  number={6315},
  pages={1024--1027},
  year={2016},
  doi={10.1126/science.aah3752}
}

@article{cong2022hardware,
  title={Hardware-efficient, fault-tolerant quantum computation with 
         {R}ydberg atoms},
  author={Cong, I. and Levine, H. and Keesling, A. and Bluvstein, D. and 
          Wang, S.-T. and Lukin, M. D.},
  journal={Physical Review X},
  volume={12},
  number={2},
  pages={021049},
  year={2022},
  doi={10.1103/PhysRevX.12.021049}
}

@article{sahay2023biased,
  title={High-threshold codes for neutral-atom qubits with biased erasure 
         errors},
  author={Sahay, K. and Jin, J. and Claes, J. and Thompson, J. D. and 
          Puri, S.},
  journal={Physical Review X},
  volume={13},
  number={4},
  pages={041013},
  year={2023},
  doi={10.1103/PhysRevX.13.041013}
}

@article{senoo2025,
  title={High-fidelity entanglement and coherent multi-qubit mapping in an atom array},
  author={Senoo, Aruku and Baumg{\"a}rtner, Alexander and Lis, Joanna W and Vaidya, Gaurav M and Zeng, Zhongda and Giudici, Giuliano and Pichler, Hannes and Kaufman, Adam M},
  journal={Nature Physics},
  volume={22},
  number={6},
  pages={903--909},
  year={2026},
  publisher={Nature Publishing Group}
}

@article{chow2024leakage,
  title={Circuit-based leakage-to-erasure conversion in a neutral-atom 
         quantum processor},
  author={Chow, M. N. H. and Buchemmavari, V. and Omanakuttan, S. and 
          Little, B. J. and Pandey, S. and Deutsch, I. H. and Jau, Y.-Y.},
  journal={PRX Quantum},
  volume={5},
  number={4},
  pages={040343},
  year={2024},
  doi={10.1103/PRXQuantum.5.040343}
}

@article{graham2023midcircuit,
  title={Midcircuit measurements on a single-species neutral alkali atom 
         quantum processor},
  author={Graham, T. M. and Phuttitarn, L. and Chinnarasu, R. and Song, Y. 
          and Poole, C. and Jooya, K. and Scott, J. and Scott, A. and 
          Eichler, P. and Saffman, M.},
  journal={Physical Review X},
  volume={13},
  number={4},
  pages={041051},
  year={2023},
  doi={10.1103/PhysRevX.13.041051}
}

@article{fowler2012surface,
  title={Surface codes: Towards practical large-scale quantum computation},
  author={Fowler, A. G. and Mariantoni, M. and Martinis, J. M. and Cleland, 
          A. N.},
  journal={Physical Review A},
  volume={86},
  number={3},
  pages={032324},
  year={2012},
  doi={10.1103/PhysRevA.86.032324}
}

@article{beverland2022assessing,
  title={Assessing requirements to scale to practical quantum advantage},
  author={Beverland, M. E. and Murali, P. and Troyer, M. and Svore, K. M. 
          and Hoefler, T. and Kliuchnikov, V. and Low, G. H. and Soeken, M. 
          and Sundaram, A. and Vaschillo, A.},
  journal={arXiv preprint arXiv:2211.07629},
  year={2022}
}

@INPROCEEDINGS {10946298,
author = { Lin, Wan-Hsuan and Tan, Daniel Bochen and Cong, Jason },
booktitle = { 2025 IEEE International Symposium on High Performance Computer Architecture (HPCA) },
title = {{ Reuse-Aware Compilation for Zoned Quantum Architectures Based on Neutral Atoms }},
year = {2025},
volume = {},
ISSN = {},
pages = {127-142},
doi = {10.1109/HPCA61900.2025.00021},
url = {https://doi.ieeecomputersociety.org/10.1109/HPCA61900.2025.00021},
publisher = {IEEE Computer Society},
address = {Los Alamitos, CA, USA},
month =mar}

@article{Graham2021MultiqubitEA,
  title={Multi-qubit entanglement and algorithms on a neutral-atom quantum computer},
  author={T. M. Graham and Y. P. Song and J. Scott and Cody Poole and L. Phuttitarn and Kais Jooya and P. Eichler and X. Jiang and A A Marra and Brandon Grinkemeyer and Minho Kwon and Matthew Ebert and J Cherek and Martin Lichtman and Matthew Gillette and J. Gilbert and David N. Bowman and Timothy G. Ballance and Catherine Campbell and Edward D. Dahl and Ophelia Crawford and Nick S. Blunt and Benjamin Rogers and Thomas W. Noel and Mark Saffman},
  journal={Nature},
  year={2021},
  volume={604},
  pages={457 - 462},
  url={https://api.semanticscholar.org/CorpusID:245537276}
}

@Article{Bluvstein2024,
author={Bluvstein, Dolev
and Evered, Simon J.
and Geim, Alexandra A.
and Li, Sophie H.
and Zhou, Hengyun
and Manovitz, Tom
and Ebadi, Sepehr
and Cain, Madelyn
and Kalinowski, Marcin
and Hangleiter, Dominik
and Bonilla Ataides, J. Pablo
and Maskara, Nishad
and Cong, Iris
and Gao, Xun
and Sales Rodriguez, Pedro
and Karolyshyn, Thomas
and Semeghini, Giulia
and Gullans, Michael J.
and Greiner, Markus
and Vuleti{\'{c}}, Vladan
and Lukin, Mikhail D.},
title={Logical quantum processor based on reconfigurable atom arrays},
journal={Nature},
year={2024},
month={Feb},
day={01},
volume={626},
number={7997},
pages={58-65},
issn={1476-4687},
doi={10.1038/s41586-023-06927-3},
url={https://doi.org/10.1038/s41586-023-06927-3}
}

@article{Bluvstein2021AQP,
  title={A quantum processor based on coherent transport of entangled atom arrays},
  author={Dolev Bluvstein and H Levine and G. Semeghini and Tout T. Wang and S Ebadi and Marcin Kalinowski and A. Keesling and Nishad Maskara and Hannes Pichler and Markus Greiner and Vladan Vuleti{\'c} and Mikhail D. Lukin},
  journal={Nature},
  year={2021},
  volume={604},
  pages={451 - 456},
  url={https://api.semanticscholar.org/CorpusID:244954259}
}

@article{Manetsch2024ATA,
  title={A tweezer array with 6,100 highly coherent atomic qubits},
  author={Hannah J. Manetsch and Gyohei Nomura and Elie Bataille and Xudong Lv and K. H. Leung and Manuel Endres},
  journal={Nature},
  year={2024},
  volume={647},
  pages={60 - 67},
  url={https://api.semanticscholar.org/CorpusID:268531555}
}

@inproceedings{10.1145/3676642.3736128,
author = {Ruan, Jixuan and Fang, Xiang and Zhang, Hezi and Li, Ang and Humble, Travis and Ding, Yufei},
title = {PowerMove: Optimizing Compilation for Neutral Atom Quantum Computers with Zoned Architecture},
year = {2025},
isbn = {9798400710803},
publisher = {Association for Computing Machinery},
address = {New York, NY, USA},
url = {https://doi.org/10.1145/3676642.3736128},
doi = {10.1145/3676642.3736128},
booktitle = {Proceedings of the 30th ACM International Conference on Architectural Support for Programming Languages and Operating Systems, Volume 3},
pages = {163–178},
numpages = {16},
location = {Rotterdam, Netherlands},
series = {ASPLOS '25}
}

@article{baranes2026leveraging,
  title = {Leveraging Qubit Loss Detection in Fault-Tolerant Quantum Algorithms},
  author = {Baranes, Gefen and Cain, Madelyn and Ataides, J. Pablo Bonilla and Bluvstein, Dolev and Sinclair, Josiah and Vuleti\ifmmode \acute{c}\else \'{c}\fi{}, Vladan and Zhou, Hengyun and Lukin, Mikhail D.},
  journal = {Phys. Rev. X},
  volume = {16},
  issue = {1},
  pages = {011002},
  numpages = {26},
  year = {2026},
  month = {Jan},
  publisher = {American Physical Society},
  doi = {10.1103/ycwc-3myc},
  url = {https://link.aps.org/doi/10.1103/ycwc-3myc}
}

@article{Kim2016SLM,
author = {Kim, Hyosub and Lee, Woojun and Lee, Han-gyeol and Jo, Hanlae and Song, Yunheung and Ahn, Jaewook},
year = {2016},
month = {01},
pages = {},
title = {In situ single-atom array synthesis by dynamic holographic optical tweezers},
volume = {7},
journal = {Nature Communications},
doi = {10.1038/ncomms13317}
}

@misc{Guo2025AOL,
author = {Guo, Zhichao and Herk, Rik and Vredenbregt, Edgar and Kokkelmans, Servaas},
year = {2025},
month = {10},
pages = {},
title = {Acousto-optic lens for 3D shuttling of atoms in a neutral atom quantum computer},
eprint={2510.09398},
archivePrefix={arXiv},
primaryClass={physics.atom-ph},
url={https://arxiv.org/abs/2510.09398},
doi = {10.48550/arXiv.2510.09398}
}

@article{66s8-jj18,
  title = {Universal Neutral-Atom Quantum Computer with Individual Optical Addressing and Nondestructive Readout},
  author = {Radnaev, A.G. and Chung, W.C. and Cole, D.C. and Mason, D. and Ballance, T.G. and Bedalov, M.J. and Belknap, D.A. and Berman, M.R. and Blakely, M. and Bloomfield, I.L. and Buttler, P.D. and Campbell, C. and Chopinaud, A. and Copenhaver, E. and Dawes, M.K. and Eubanks, S.Y. and Friss, A.J. and Garcia, D.M. and Gilbert, J. and Gillette, M. and Goiporia, P. and Gokhale, P. and Goldwin, J. and Goodwin, D. and Graham, T.M. and Guttormsson, C.J. and Hickman, G.T. and Hurtley, L. and Iliev, M. and Jones, E.B. and Jones, R.A. and Kuper, K.W. and Lewis, T.B. and Lichtman, M.T. and Majdeteimouri, F. and Mason, J.J. and McMaster, J.K. and Miles, J.A. and Mitchell, P.T. and Murphree, J.D. and Neff-Mallon, N.A. and Oh, T. and Omole, V. and Parlo Simon, C. and Pederson, N. and Perlin, M.A. and Reiter, A. and Rines, R. and Romlow, P. and Scott, A.M. and Stiefvater, D. and Tanner, J.R. and Tucker, A.K. and Vinogradov, I.V. and Warter, M.L. and Yeo, M. and Saffman, M. and Noel, T.W.},
  journal = {PRX Quantum},
  volume = {6},
  issue = {3},
  pages = {030334},
  numpages = {20},
  year = {2025},
  month = {Aug},
  publisher = {American Physical Society},
  doi = {10.1103/66s8-jj18},
  url = {https://link.aps.org/doi/10.1103/66s8-jj18}
}

@Article{Evered2023,
author={Evered, Simon J.
and Bluvstein, Dolev
and Kalinowski, Marcin
and Ebadi, Sepehr
and Manovitz, Tom
and Zhou, Hengyun
and Li, Sophie H.
and Geim, Alexandra A.
and Wang, Tout T.
and Maskara, Nishad
and Levine, Harry
and Semeghini, Giulia
and Greiner, Markus
and Vuleti{\'{c}}, Vladan
and Lukin, Mikhail D.},
title={High-fidelity parallel entangling gates on a neutral-atom quantum computer},
journal={Nature},
year={2023},
month={Oct},
day={01},
volume={622},
number={7982},
pages={268-272},
issn={1476-4687},
doi={10.1038/s41586-023-06481-y},
url={https://doi.org/10.1038/s41586-023-06481-y}
}

@misc{Evered2026,
      title={High-fidelity entangling gates and nonlocal circuits with neutral atoms}, 
      author={Simon J. Evered and Muqing Xu and Sophie H. Li and Alexandra A. Geim and J. Pablo Bonilla Ataides and Marcin Kalinowski and Dolev Bluvstein and Nishad Maskara and Christian Kokail and Markus Greiner and Vladan Vuletić and Mikhail D. Lukin},
      year={2026},
      eprint={2604.25987},
      archivePrefix={arXiv},
      primaryClass={quant-ph},
      url={https://arxiv.org/abs/2604.25987}, 
}

@misc{computing2026quantumerrorcorrectiontoric,
      title={Quantum error correction with the toric code}, 
      author={Atom and Computing and Collaborators},
      year={2026},
      eprint={2606.04079},
      archivePrefix={arXiv},
      primaryClass={quant-ph},
      url={https://arxiv.org/abs/2606.04079}, 
}

@article{grimm2000optical,
  title={Optical dipole traps for neutral atoms},
  author={Grimm, R. and Weidem{\"u}ller, M. and Ovchinnikov, Y. B.},
  journal={Advances in Atomic, Molecular, and Optical Physics},
  volume={42},
  pages={95--170},
  year={2000},
  doi={10.1016/S1049-250X(08)60186-X}
}

@article{sahay2025fold,
  title = {Fold-transversal surface code cultivation},
  author = {Sahay, Kaavya and Tsai, Pei-Kai and Chang, Kathleen (Katie) and Su, Qile and Smith, Thomas B. and Singh, Shraddha and Puri, Shruti},
  journal = {PRX Quantum},
  volume = {7},
  issue = {3},
  pages = {033006},
  numpages = {26},
  year = {2026},
  month = {Jul},
  publisher = {American Physical Society},
  doi = {10.1103/gpvl-lg4c},
  url = {https://link.aps.org/doi/10.1103/gpvl-lg4c}
}

@article{derks2025designing,
  title={Designing fault-tolerant circuits using detector error models},
  author={Derks, Peter-Jan HS and Townsend-Teague, Alex and Burchards, Ansgar G and Eisert, Jens},
  journal={Quantum},
  volume={9},
  pages={1905},
  year={2025},
  publisher={Verein zur F{\"o}rderung des Open Access Publizierens in den Quantenwissenschaften}
}

@article{gidney2024magic,
  title={Magic state cultivation: growing T states as cheap as CNOT gates},
  author={Gidney, Craig and Shutty, Noah and Jones, Cody},
  journal={arXiv preprint arXiv:2409.17595},
  year={2024}
}

@article{jacoby2025magic,
  title={Magic state injection with erasure qubits},
  author={Jacoby, Shoham and Vaknin, Yotam and Retzker, Alex and Grimsmo, Arne L},
  journal={PRX Quantum},
  volume={6},
  number={4},
  pages={040323},
  year={2025},
  publisher={APS}
}

@misc{ross2016optimalancillafreecliffordtapproximation,
      title={Optimal ancilla-free Clifford+T approximation of z-rotations}, 
      author={Neil J. Ross and Peter Selinger},
      year={2016},
      eprint={1403.2975},
      archivePrefix={arXiv},
      primaryClass={quant-ph},
      url={https://arxiv.org/abs/1403.2975}, 
}

@misc{obenland_2026_18154991,
  author       = {Obenland, Kevin and
                  Elenewski, Justin and
                  Morrell, Kaitlyn and
                  Rempfer, Benjamin and
                  Kuklinski, Parker and
                  Neumann, Rylee Stuart and
                  Kurlej, Arthur and
                  Rood, Robert and
                  Blue, John and
                  Belarge, Joe},
  title        = {pyLIQTR},
  month        = jan,
  year         = 2026,
  publisher    = {Zenodo},
  version      = {v1.4.2},
  doi          = {10.5281/zenodo.18154991},
  url          = {https://doi.org/10.5281/zenodo.18154991},
  swhid        = {swh:1:dir:256a5a3b521277cda43bbbc7d38c5d1d114f0e7a
                   ;origin=https://doi.org/10.5281/zenodo.7221272;vis
                   it=swh:1:snp:9eb6cc3f6f3526c0eceda9a6055a25a1c78ec
                   21a;anchor=swh:1:rel:ade2d21d0309518ee0ec6d422b82d
                   0fb2ebc4ac2;path=isi-usc-edu-pyLIQTR-0465e1d
                  },
}

@article{Hwang:25,
author = {Sunhwa Hwang and Hansub Hwang and Kangjin Kim and Andrew Byun and Kangheun Kim and Seokho Jeong and Maynardo Pratama Soegianto and Jaewook Ahn},
journal = {Optica Quantum},
number = {1},
pages = {64--71},
publisher = {Optica Publishing Group},
title = {Fast and reliable atom transport by optical tweezers},
volume = {3},
month = {Feb},
year = {2025},
url = {https://opg.optica.org/opticaq/abstract.cfm?URI=opticaq-3-1-64},
doi = {10.1364/OPTICAQ.546797},
}

@article{j2fw-ccmy,
  title = {Transversal Architecture for Megaquop-Scale Quantum Simulation with Neutral Atoms},
  author = {Ismail, Refaat and Chen, I-Chi and Zhao, Chen and Weiss, Ronen and Liu, Fangli and Zhou, Hengyun and Wang, Sheng-Tao and Sornborger, Andrew and Kornja\ifmmode \check{c}\else \v{c}\fi{}a, Milan},
  journal = {PRX Quantum},
  volume = {7},
  issue = {2},
  pages = {020343},
  numpages = {39},
  year = {2026},
  month = {May},
  publisher = {American Physical Society},
  doi = {10.1103/j2fw-ccmy},
  url = {https://link.aps.org/doi/10.1103/j2fw-ccmy}
}

@article{gidney2025yoked,
  title={Yoked surface codes},
  author={Gidney, Craig and Newman, Michael and Brooks, Peter and Jones, Cody},
  journal={Nature Communications},
  volume={16},
  number={1},
  pages={4498},
  year={2025},
  publisher={Nature Publishing Group UK London}
}

@article{Higgott2025sparseblossom,
  doi = {10.22331/q-2025-01-20-1600},
  url = {https://doi.org/10.22331/q-2025-01-20-1600},
  title = {Sparse {B}lossom: correcting a million errors per core second with minimum-weight matching},
  author = {Higgott, Oscar and Gidney, Craig},
  journal = {{Quantum}},
  issn = {2521-327X},
  publisher = {{Verein zur F{\"{o}}rderung des Open Access Publizierens in den Quantenwissenschaften}},
  volume = {9},
  pages = {1600},
  month = jan,
  year = {2025}
}

@article{perrin2025quantum,
  title={Quantum error correction resilient against atom loss},
  author={Perrin, Hugo and Jandura, Sven and Pupillo, Guido},
  journal={Quantum},
  volume={9},
  pages={1884},
  year={2025},
  publisher={Verein zur F{\"o}rderung des Open Access Publizierens in den Quantenwissenschaften}
}

@inproceedings{10.1109/ISCA59077.2024.00030,
author = {Wang, Hanrui and Liu, Pengyu and Tan, Daniel Bochen and Liu, Yilian and Gu, Jiaqi and Pan, David Z. and Cong, Jason and Acar, Umut A. and Han, Song},
title = {Atomique: A Quantum Compiler for Reconfigurable Neutral Atom Arrays},
year = {2025},
isbn = {9798350326581},
publisher = {IEEE Press},
url = {https://doi.org/10.1109/ISCA59077.2024.00030},
doi = {10.1109/ISCA59077.2024.00030},
booktitle = {Proceedings of the 51st Annual International Symposium on Computer Architecture},
pages = {293–309},
numpages = {17},
location = {Buenos Aires, Argentina},
series = {ISCA '24}
}

@article{rodriguez2025experimental,
  title = {Experimental demonstration of logical magic state distillation},
  author = {Sales Rodriguez, Pedro and Robinson, John M. and Jepsen, Paul Niklas and He, Zhiyang and Duckering, Casey and Zhao, Chen and Wu, Kai-Hsin and Campo, Joseph and Bagnall, Kevin and Kwon, Minho and others},
  journal = {Nature},
  volume = {645},
  pages = {620--625},
  year = {2025},
  doi = {10.1038/s41586-025-09367-3},
  url = {https://doi.org/10.1038/s41586-025-09367-3}
}

@article{chiu2025continuous,
  title = {Continuous operation of a coherent 3,000-qubit system},
  author = {Chiu, Neng-Chun and Trapp, Elias C. and Guo, Jinen and Abobeih, Mohamed H. and Stewart, Luke M. and Hollerith, Simon and Stroganov, Pavel and Kalinowski, Marcin and Geim, Alexandra A. and Evered, Simon J. and Li, Sophie H. and Peters, Lisa M. and Bluvstein, Dolev and Wang, Tout T. and Greiner, Markus and Vuleti{\'c}, Vladan and Lukin, Mikhail D.},
  journal = {Nature},
  volume = {646},
  pages = {1075--1080},
  year = {2025},
  doi = {10.1038/s41586-025-09596-6},
  url = {https://doi.org/10.1038/s41586-025-09596-6}
}

@article{norcia2024iterative,
	title = {Iterative {Assembly} of ${}^{171}\mathrm{Yb}$ {Atom} {Arrays} with {Cavity}-{Enhanced} {Optical} {Lattices}},
	volume = {5},
	url = {https://link.aps.org/doi/10.1103/PRXQuantum.5.030316},
	doi = {10.1103/PRXQuantum.5.030316},
	number = {3},
	journal = {PRX Quantum},
	author = {Norcia, M. A. and Kim, H. and Cairncross, W. B. and Stone, M. and Ryou, A. and Jaffe, M. and Brown, M. O. and Barnes, K. and Battaglino, P. and Bohdanowicz, T. C. and Brown, A. and Cassella, K. and Chen, C.-A. and Coxe, R. and Crow, D. and others},
	year = {2024},
	pages = {030316},
}

@article{suchara2015leakage,
author = {Suchara, Martin and Cross, Andrew W. and Gambetta, Jay M.},
title = {Leakage suppression in the Toric code},
year = {2015},
issue_date = {September 2015},
publisher = {Rinton Press, Incorporated},
address = {Paramus, NJ},
volume = {15},
number = {11–12},
issn = {1533-7146},
journal = {Quantum Info. Comput.},
month = sep,
pages = {997–1016},
numpages = {20}
}

@inproceedings{Li2025FastCA,
  title={Fast, continuous and coherent atom replacement in a neutral atom qubit array},
  author={Yiyi Li and Yicheng Bao and Michael Peper and Chenyuan Li and Jeff D. Thompson},
  year={2025},
  booktitle={https://api.semanticscholar.org/CorpusID:279447782}
}

@article{gidney2021stim,
  doi = {10.22331/q-2021-07-06-497},
  url = {https://doi.org/10.22331/q-2021-07-06-497},
  title = {Stim: a fast stabilizer circuit simulator},
  author = {Gidney, Craig},
  journal = {{Quantum}},
  issn = {2521-327X},
  publisher = {{Verein zur F{\"{o}}rderung des Open Access Publizierens
                in den Quantenwissenschaften}},
  volume = 5,
  pages = 497,
  month = jul,
  year = 2021
}

@article{li2022qasmbench,
  title={QASMBench: A Low-Level Quantum Benchmark Suite for NISQ Evaluation and Simulation},
  author={Li, Ang and Stein, Samuel and Krishnamoorthy, Sriram and Ang, James},
  journal={ACM Transactions on Quantum Computing},
  year={2022},
  publisher={ACM New York, NY}
}

@article{cain2026shor,
  title   = {Shor's algorithm is possible with as few as 10,000
             reconfigurable atomic qubits},
  author  = {Cain, Madelyn and Xu, Qian and King, Robbie and
             Picard, Lewis R. B. and Levine, Harry and Endres, Manuel and
             Preskill, John and Huang, Hsin-Yuan and Bluvstein, Dolev},
  journal = {arXiv preprint arXiv:2603.28627},
  year    = {2026}
}
